\documentclass[iop]{emulateapj}

\usepackage{amsmath,amssymb}
\usepackage{graphicx}
\usepackage{subfigure}
\usepackage{multirow}

\def\msun{\,{\rm M}_\odot}
\def\eft{\times10^{52}}
\def\ergs{\,{\rm erg}\,{\rm s}^{-1}}
\def\km{\,{\rm km}}
\def\ms{\,{\rm ms}}
\def\lnu{L_{\nu_e}}

\slugcomment{Submitted to the Astrophysical Journal}

\shorttitle{Dimensionality and the Hydrodynamics of CCSNe}
\shortauthors{Dolence et al.}

\begin{document}

\title{Dimensional Dependence of the Hydrodynamics of Core-Collapse Supernovae}
\author{Joshua C. Dolence\altaffilmark{1}, Adam Burrows\altaffilmark{1}, Jeremiah W. Murphy\altaffilmark{1}, Jason Nordhaus\altaffilmark{2}}
\altaffiltext{1}{Department of Astrophysical Sciences, Princeton University, Princeton, NJ 08544}
\altaffiltext{2}{NSF Fellow, Center for Computational Relativity and Gravitation, Rochester Institute of Technology, Rochester, NY 14623}
\email{jdolence@astro.princeton.edu, burrows@astro.princeton.edu,}
\email{jmurphy@astro.princeton.edu, nordhaus@astro.rit.edu}

\begin{abstract}
The multidimensional character of the hydrodynamics in core-collapse supernova (CCSN) cores is a key facilitator of explosions.  A major goal over the last decade has been to elucidate how multidimensionality produces more favorable conditions for explosions.  Unfortunately, much of this work has necessarily been performed assuming axisymmetry and it remains unclear whether or not this compromises those results.  In this work, we present analyses of simplified two- and three-dimensional CCSN models with the goal of comparing the multidimensional hydrodynamics in setups that differ only in dimension.  Not surprisingly, we find many differences between 2D and 3D models.  While some differences are subtle and perhaps not crucial to understanding the explosion mechanism, others are quite dramatic and make interpreting 2D CCSN models problematic.  In particular, we find that imposing axisymmetry artificially produces excess power at the largest spatial scales, power that has been deemed critical in the success of previous explosion models and has been attributed solely to the standing accretion shock instability.  Nevertheless, our 3D models, which have an order of magnitude less power on large scales compared to 2D models, explode earlier.   Since we see explosions earlier in 3D than in 2D, the vigorous sloshing associated with the large scale power in 2D models is either not critical in any dimension or the explosion mechanism operates differently in 2D and 3D.  Possibly related to the earlier explosions in 3D, we find that about 25\% of the accreted material spends more time in the gain region in 3D than in 2D, being exposed to more integrated heating and reaching higher peak entropies, an effect we associate with the differing characters of turbulence in 2D and 3D.  Finally, we discuss a simple model for the runaway growth of buoyant bubbles that is able to quantitatively account for the growth of the shock radius and predicts a critical luminosity relation.
\end{abstract}


\section{Introduction}
Multidimensional hydrodynamic instabilities play a central role in the explosion mechanism of most core-collapse supernovae (CCSNe).  Though this has been repeatedly demonstrated in a variety of contexts \citep[][]{Hera92,Burr93,Hera94,Burr95a,Jank96,Mare09}, there is as yet no definitive understanding of the role of multidimensional effects in facilitating explosions.  Important effects may include increased dwell times in the gain region \citep[][]{Murp08}, expansion of the shock due to turbulent pressure support from neutrino-driven convection \citep[][]{Hera94,Burr95a,Murp12} and/or the standing accretion shock instability (SASI) \citep[e.g.][]{Sche08}, simultaneous accretion and explosion \citep[][]{Burr95a}, suppression of cooling beneath the gain region \citep{Pejc12}, and other still unidentified processes.

With the exception of a few preliminary results in 3D \citep{Brue09,Kuro12,Taki12}, the enormous computational expense has limited the most sophisticated supernova models, including the multi-species, multi-group neutrino transport, to 2D axisymmetric simulations \citep{Ott08,Mull12}.  Unfortunately, the fundamentally 3D hydrodynamics in the post-shock turbulent flow is qualitatively different if axisymmetry is imposed, as we discuss in detail below.  Understanding the differences between 2D and 3D behavior and how dimension effects the mechanism of explosion is critical in the interpretation of realistic 2D simulations and ultimately in elucidating how real stars explode.  If we find explosions in axisymmetry, should we expect them in 3D?  If so, are the conditions identified in 2D as crucial to producing explosions manifest in 3D?  How are explosions triggered in 3D?

In this work, we do not perform sophisticated radiation hydrodynamic simulations of realistic core-collapse supernovae.  Rather, we perform a series of simplified numerical experiments designed to clarify how dimension effects the hydrodynamics leading to explosions \citep[see also][]{Murp08,Nord10,Hank12}.  We find that there are many differences between 2D and 3D models, some quite dramatic, and that conditions identified as important in 2D may not remain so in more realistic 3D models.  Nevertheless, our 3D models explode whenever the 2D models explode and, moreover, 3D models explode earlier.

We begin in Sec.~\ref{sec:setup} with a discussion of our numerical setup and solution technique.  Section~\ref{sec:overview} gives an overview of the basic results of our simulations and discusses qualitative differences in the structures of 2D and 3D models.  Section~\ref{sec:global} begins the quantitative analyses of the simulations, showing how the global structures of the flows are different between 2D and 3D models.  Section~\ref{sec:turb} compares various measures of the 2D and 3D turbulence in the post-shock flows.  Section~\ref{sec:expl} discusses popular explosion metrics and introduces a simple model for explosions based on the runaway growth of bubbles.  Finally, Sec.~\ref{sec:conc} discusses our results and conclusions.

\section{Numerical Setup}\label{sec:setup}
Our setup is the same as that presented in \citet{Burr12} and \citet{Murp12}.  We use the CASTRO adaptive mesh refinement hydrodynamics code \citep{Almg10} to evolve two-dimensional (axisymmetric) and three-dimensional models of the collapse, bounce, and subsequent evolution of the $15$-$\msun$ non-rotating solar-metallicity progenitor of \citet{Woos95}.  The adaptive mesh uses six levels of factor two refinement with $\approx0.5\km$ resolution in the inner $50\km$ and $2\km$ or better resolution everywhere behind and including the shock during the stalled pre-explosion phase.  We ensure that both two- and three-dimensional simulations utilize the same refinement criteria to minimize any differences that might arise from different grid structures.  The domains include the inner $5000\km$ in radius.

As in previous works, we adopt the ``light bulb'' prescription for neutrino heating and cooling \citep{Murp08,Nord10,Hank12,Burr12,Murp12,Couc12}.  In this prescription, neutrino heating is parametrized by a constant driving electron neutrino luminosity $\lnu$ (the electron neutrino and antineutrino luminosities are assumed to be equal) and we present results for three luminosities: $2.1\eft\ergs$, $2.2\eft\ergs$, and $2.3\eft\ergs$.  Neutrino cooling occurs at a rate $\propto T^6$ \citep[][]{Beth90}.  Since we do not explicitly treat the neutrino transport, the electron fraction $Y_e$ is evolved according to the prescription given by \citet{Lieb05}.  Finally, the equation of state is based on the relativistic mean-field theory of \citet{Shen98a,Shen98b} and we assume nuclear statistical equilibrium.

\section{Overview of Simulation Results}\label{sec:overview}
We find that the structure and evolution of the post-bounce hydrodynamics can be clearly distinguished between simulations that differ only in dimension.  In this section, we develop a qualitative view of the structure and evolution of 2D and 3D models.

\begin{figure}[ht]
\centering
\includegraphics[width=\columnwidth]{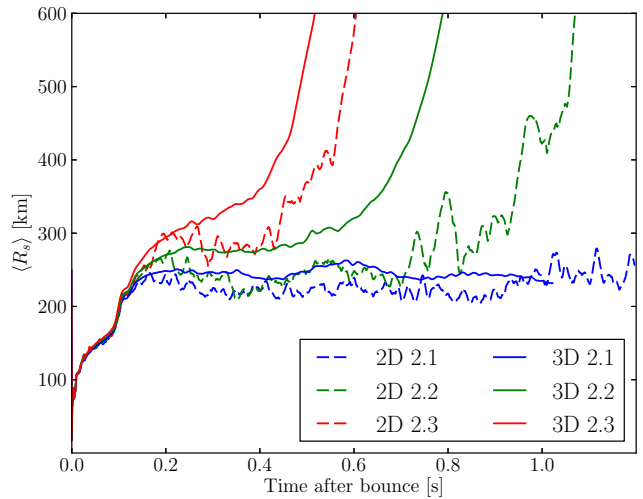}
\caption{Average shock radii for all six models considered in this work.  The very early phases are quite similar, but the models diverge after $100\ms$ post-bounce.  In the quasi-steady accretion phase, the average shock radius is $\sim$$30\km$ larger in 3D (solid) than in 2D (dashed) for a given driving neutrino luminosity.  Finally, when explosions occur, they set in $\sim$$100$--$300\ms$ earlier in 3D than in 2D.}
\label{fig:rshock}
\end{figure}

Figure~\ref{fig:rshock} shows the development of the average shock radius for all three driving neutrino luminosities in both 2D and 3D.  For the first $100\ms$ after bounce, the models are nearly indistinguishable, but diverge thereafter.  The stalled shock radii in 3D are generally larger than in 2D and are less variable in both angle and time.  The models with driving luminosities of $2.2\eft\ergs$ and $2.3\eft\ergs$ have sufficient neutrino heating to explode within the simulated time in both 2D and 3D.  The explosions, however, occur earlier in 3D than in 2D, in qualitative agreement with \citet{Nord10} despite deficiencies in that work \citep{Burr12}.  Moreover, once explosions set in, the shock radii grow monotonically in 3D whereas they show oscillatory behavior in 2D, at least early in the explosion phase.

\begin{figure}[htb]
\centering
\includegraphics[width=\columnwidth]{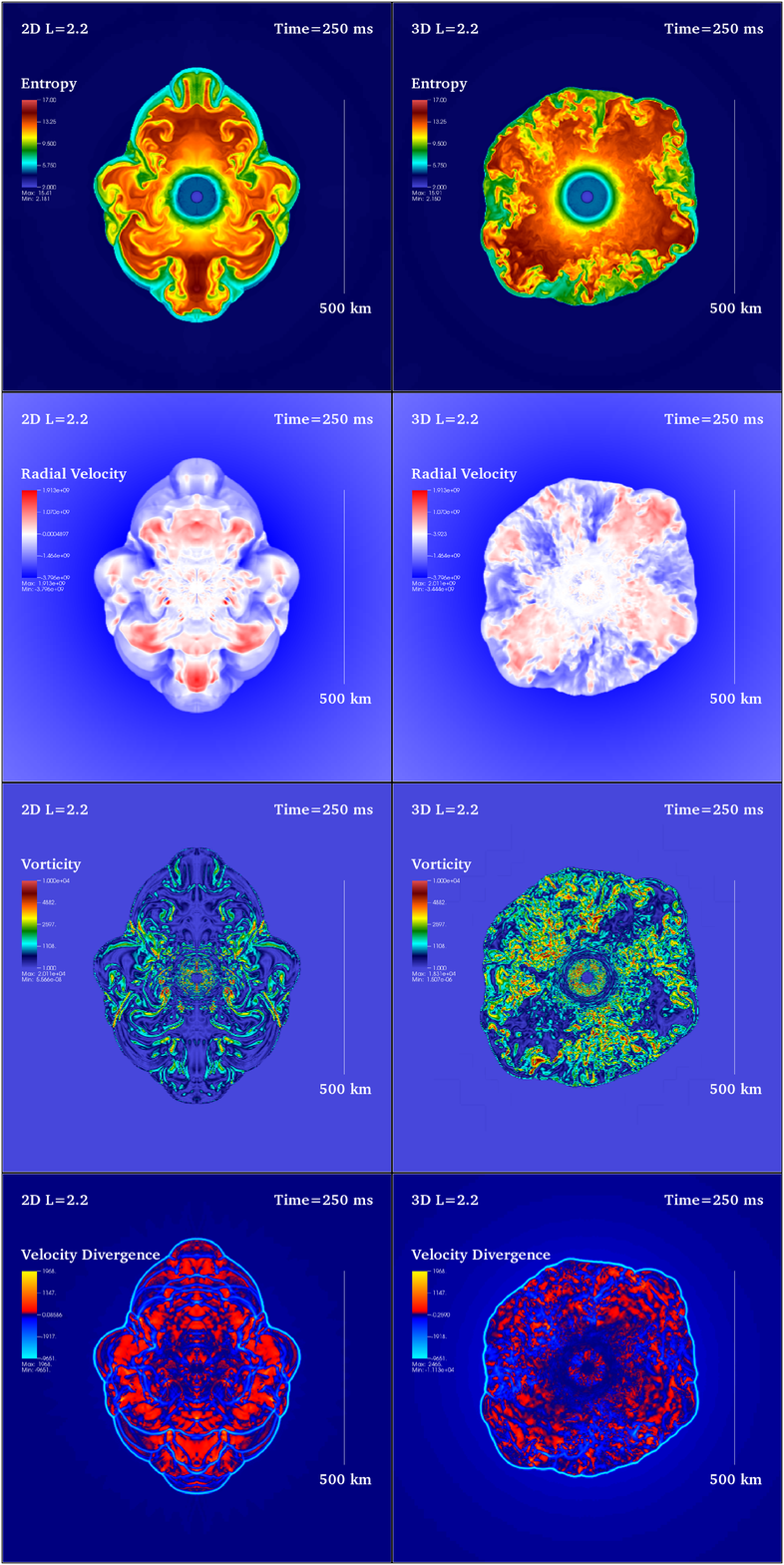}
\caption{Snapshots of the entropy, radial velocity, magnitude of the vorticity ($|\nabla\times\vec{v}|$), and velocity divergence ($\nabla\cdot\vec{v}$) at $250\ms$ post-bounce for the 2D (left) and 3D (slice, right) $\lnu=2.2\eft\ergs$ models.  As evident in all four quantities and in contrast with the 2D models, 3D models show significantly more small scale structure, have no preferred axis, and tend to be more spherical in the early stalled accretion shock phase.}
\label{fig:snap250}
\end{figure}
\begin{figure}[htb]
\centering
\includegraphics[width=\columnwidth]{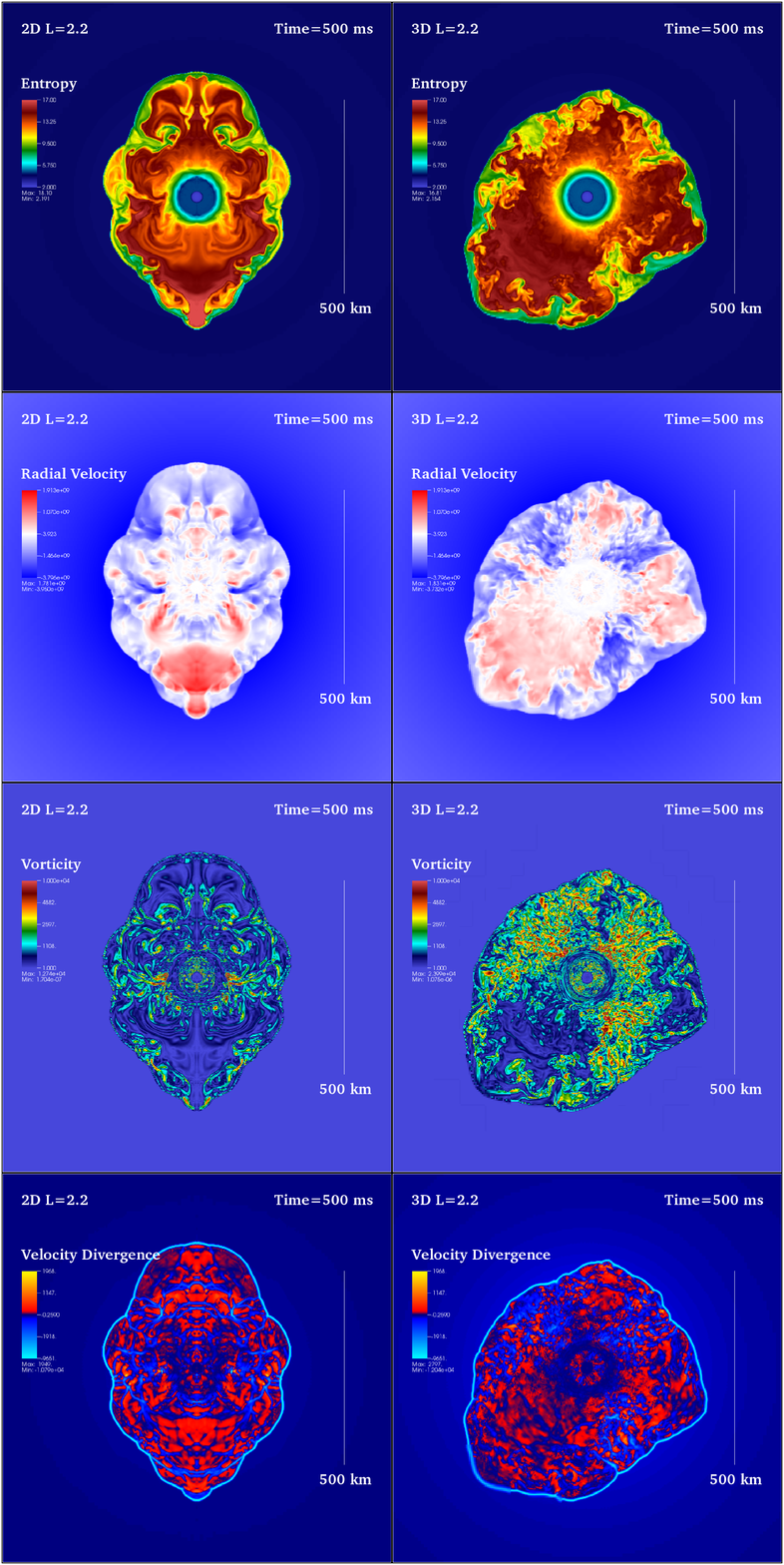}
\caption{Same as Fig.~\ref{fig:snap250}, but at $500\ms$ post-bounce.  As compared with the snapshot at $250\ms$, the 3D model is developing significant asymmetry as it evolves towards explosion.  The 2D model is qualitatively similar at $500\ms$ post-bounce to $250\ms$ post-bounce, with its distinctive prolate distortion and characteristically larger structures compared with the 3D model.}
\label{fig:snap500}
\end{figure}

To develop a qualitative sense for some of the underlying hydrodynamic differences, we show snapshots of various quantities at select times in the evolution.  Figure~\ref{fig:snap250} shows slices of the entropy, radial velocity, magnitude of the vorticity ($|\nabla\times\vec{v}|$), and velocity divergence ($\nabla\cdot\vec{v}$) at $250\ms$ post-bounce for the $\lnu=2.2\eft\ergs$ models in 2D and 3D.  Figure~\ref{fig:snap500} shows the same quantities at $500\ms$ post-bounce.  Perhaps the most obvious difference between 2D and 3D is the clear presence of a preferred axis, leading to a distinctly prolate distortion of the post-shock flow.  This is a generic result seen in all 2D simulations, even those that include sophisticated neutrino transport and relativistic effects \citep[][]{Ott08,Mare09,Mull12}.  Importantly, this may be an \emph{artifact} of assuming axisymmetry, the consequences of which are difficult to clarify.  Another important difference, identified previously, is the existence of more small-scale structure in the flow in 3D \citep{Hank12}, as can be seen in the entropy, radial velocity, vorticity, and velocity divergence.  Interestingly, when comparing the entropy and radial velocity maps in both the 2D and 3D models, there is a strong correspondence between large-scale structures; high-entropy plumes are associated with outflow whereas low-entropy regions are associated with inflow, a natural consequence of buoyancy-driven convection, which it has been argued dominates the flow in the stalled accretion shock phase \citep{Burr12,Murp12}.  In 3D, these rising plumes also have lower vorticities and larger velocity divergences than the surrounding flow, suggesting that these structures are relatively coherent and expanding.  In 2D, these associations are more difficult to make by eye.  

The 2D and 3D post-shock flows have qualitative differences that are easily identified.  The geometries of the flows are different and the characteristics of the turbulent, convective, post-shock flows are different.  One consequence of these differences is a larger averaged stalled shock radius in 3D relative to 2D.  In the remainder of this work, we undertake detailed analyses of the post-shock hydrodynamics to better understand how these qualitative differences manifest quantitatively in quantities that have been suggested to be of importance.

\section{Global Structures}\label{sec:global}
\subsection{Integral Quantities}
\begin{figure}[htb]
\centering
\includegraphics[width=\columnwidth]{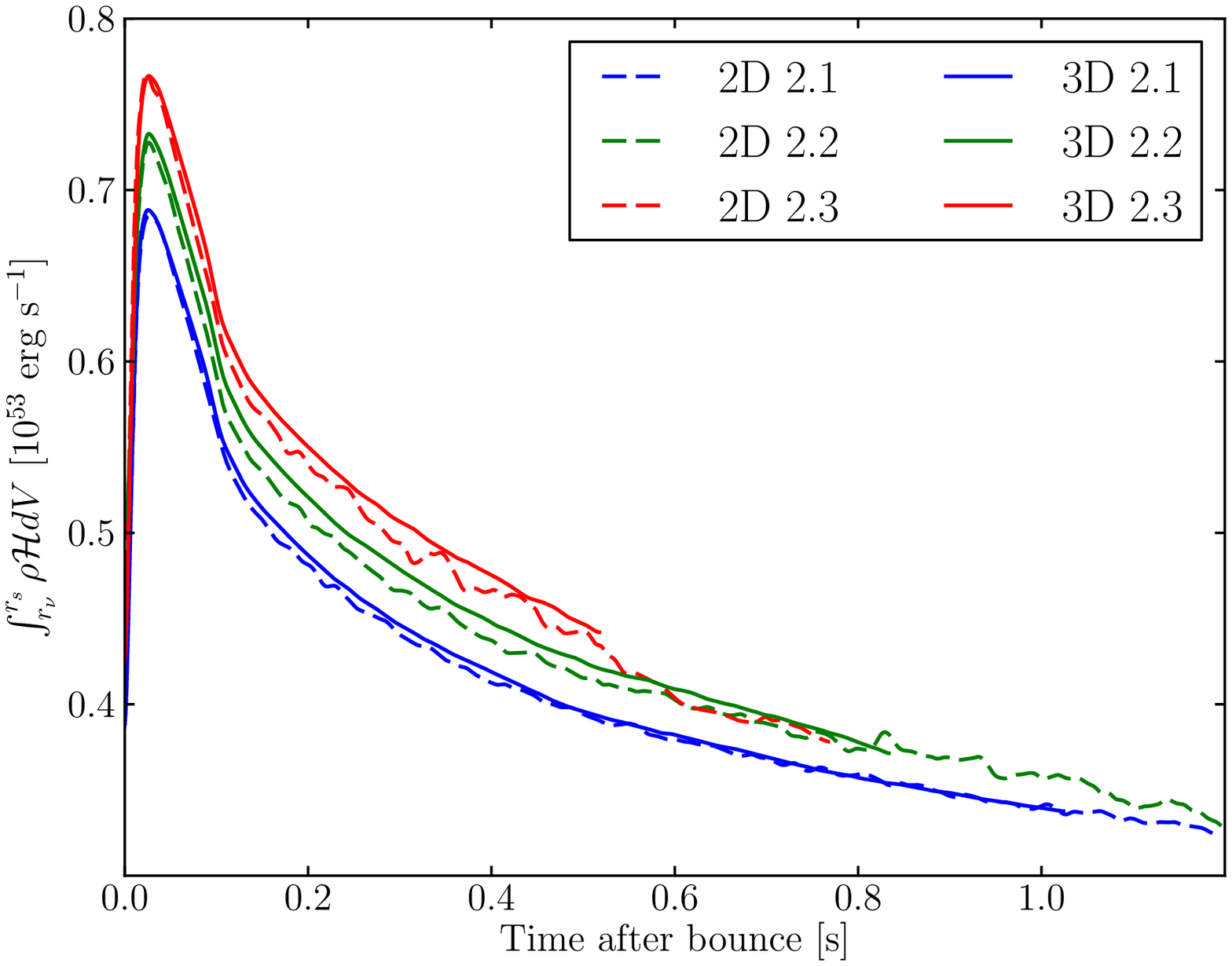}
\caption{Integrated rate of neutrino heating in the volume between the neutrinosphere ($r_\nu$) and shock ($r_s$).  The rates are quite comparable between 2D (dashed) and 3D (solid), but 3D tends to be a few percent ($\sim$$1$--$3\%$) higher.}
\label{fig:total_heat}
\end{figure}

\begin{figure}[htb]
\centering
\includegraphics[width=\columnwidth]{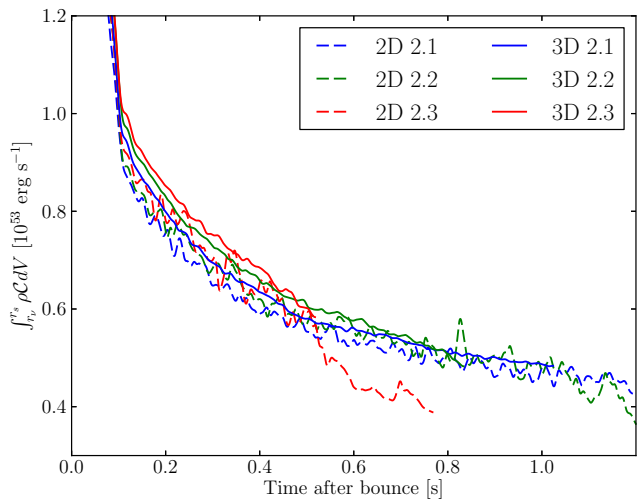}
\caption{Integrated rate of neutrino cooling in the volume between the neutrinosphere ($r_\nu$) and shock ($r_s$).  The 3D (solid) models tend to cool more ($\sim$$10\%$) than their 2D (dashed) counterparts.}
\label{fig:total_cool}
\end{figure}

\begin{figure}[htb]
\centering
\includegraphics[width=\columnwidth]{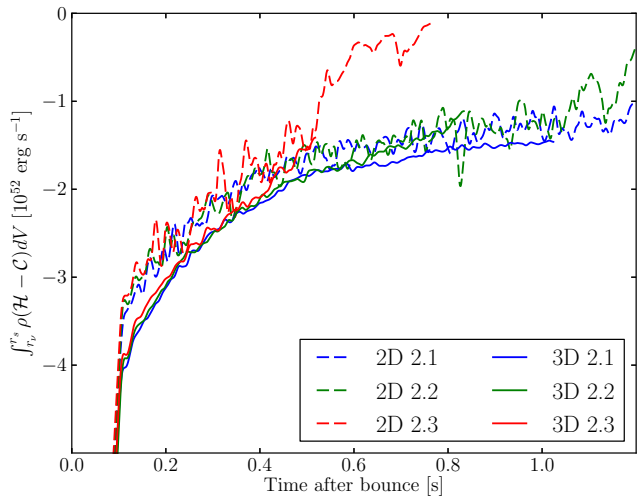}
\caption{Net rate of neutrino heating in the volume between the neutrinosphere ($r_\nu$) and shock ($r_s$).  Note that the heating rates are all negative, indicating that there is net neutrino cooling overall.  In spite of the marginally higher heating rate, the 3D models (solid) show less ($\sim$$10\%$) \emph{net} heating per unit time than the 2D models (dashed).}
\label{fig:total_net}
\end{figure}

\begin{figure}[htb]
\centering
\includegraphics[width=\columnwidth]{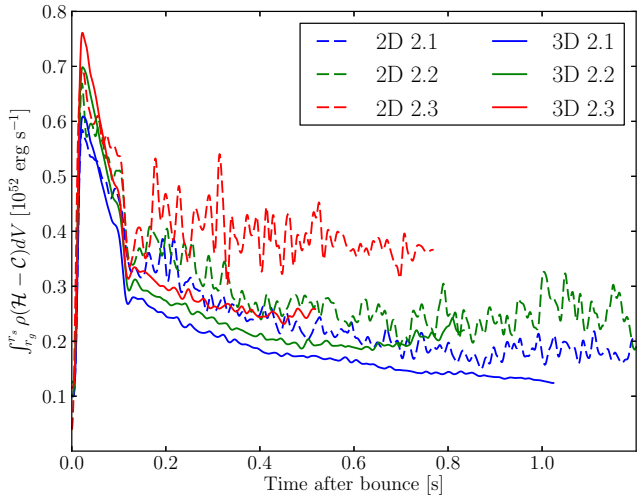}
\caption{Net neutrino heating rate in the gain region between the gain radius ($r_g$) and the shock ($r_s$).  After an initial transient phase lasting $\sim$$50\ms$, 2D models (dashed) show $\sim$$30\%$ higher heating rates in this region than the 3D models (solid) for a given driving luminosity.  The specific heating rates, found by dividing the heating rates shown above by the mass in the gain region, are also larger in 2D than in 3D.}
\label{fig:gain_net}
\end{figure}

\begin{figure}
\centering
\includegraphics[width=\columnwidth]{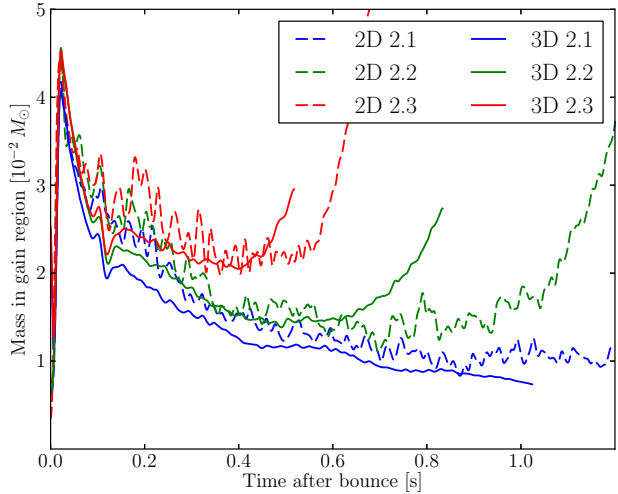}
\caption{Mass in the gain region for all six models in units of $10^{-2}\msun$.  Prior to explosion, the gain mass is generally larger in 2D (dashed) than in 3D (solid), which may contribute to the higher neutrino heating rates.}
\label{fig:mgain}
\end{figure}

\begin{figure}
\centering
\includegraphics[width=\columnwidth]{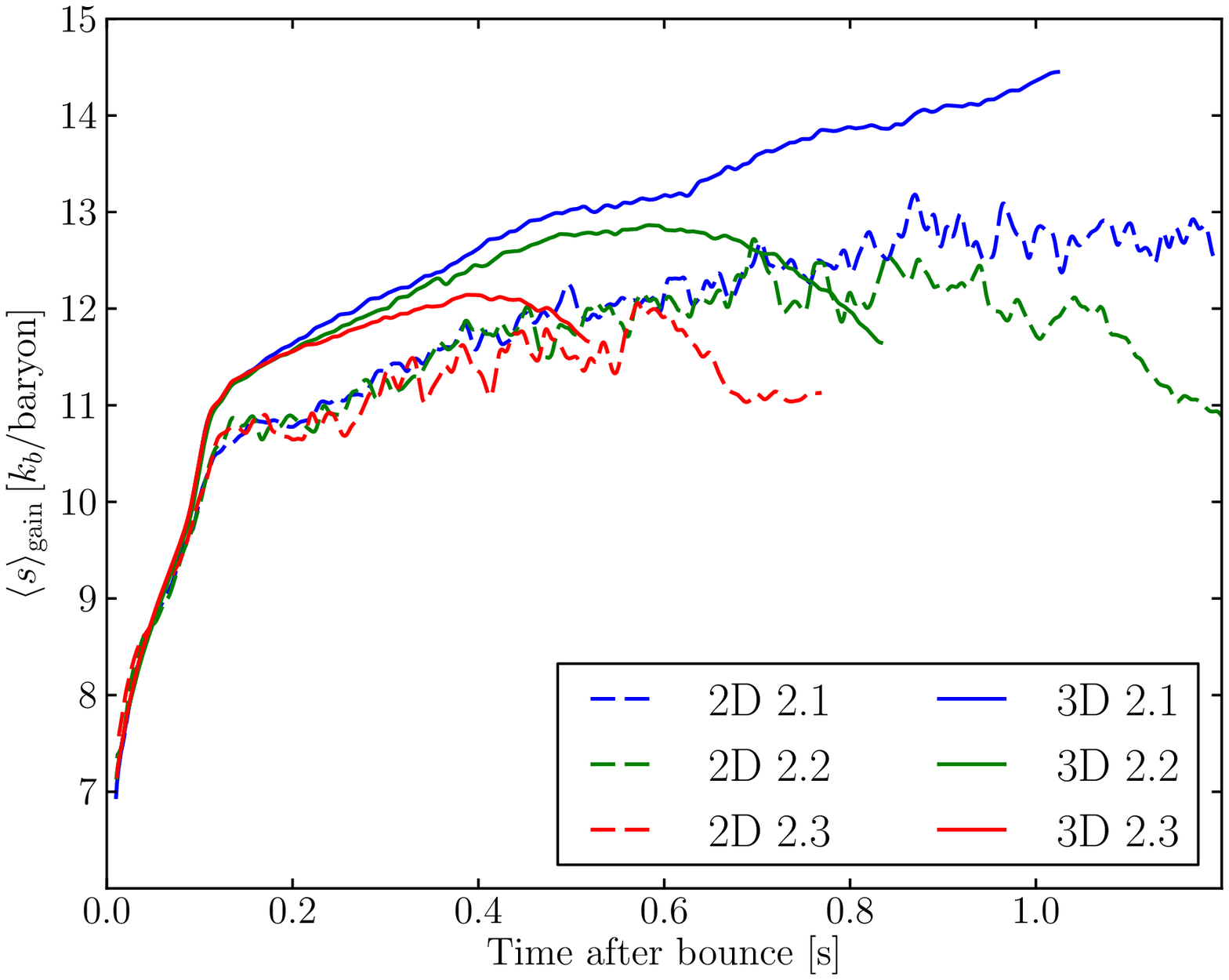}
\caption{Average specific entropy in units of $k_b/{\rm baryon}$ in the gain region.  In the stalled shock phase, the average entropies are higher in 3D (solid) by about one unit compared to 2D (dashed).  The turndown in the $\lnu=2.2\eft\ergs$ and $\lnu=2.3\eft\ergs$ models is associated with the relatively lower entropy, unheated material that enters into the post-shock region in the exploding phase.}
\label{fig:avgent}
\end{figure}

While unable to capture the full scope of the differences between 2D and 3D flows, integral quantities offer the advantages of simplicity and widespread usage.  In the context of the neutrino mechanism, the integrated heating and cooling in the post-shock flow are naturally important quantities.  Figures~\ref{fig:total_heat}, \ref{fig:total_cool}, and \ref{fig:total_net} show the total integrated heating, cooling, and net heating minus cooling rates, respectively, in the region between the neutrinosphere (where the optical depth to neutrinos is approximately unity) and shock.  While there tends to be marginally more heating in the 3D models, they have significantly more cooling and, therefore, less net heating than their corresponding 2D models.  If we focus on the gain region (i.e. only the region with net neutrino heating), we see in Fig.~\ref{fig:gain_net} that the 2D models have significantly more heating than their 3D counterparts.  This arises, in part, because, prior to explosion, there is more mass in the gain region in 2D than in 3D, as shown in Fig.~\ref{fig:mgain}.  Finally, in spite of the higher heating rates in 2D, the average specific entropy in the gain region,
\begin{equation}
\langle s \rangle_{\rm gain} = \frac{1}{M_{\rm gain}}\int_{\rm gain} \rho s dV,
\end{equation}
is larger in 3D than in 2D, as shown in Fig.~\ref{fig:avgent} and suggested by \citet{Nord10}.  This is consistent with the results shown in \citet{Hank12} who found somewhat larger average entropies in 3D than in 2D.  We note, however, that a higher average entropy is not directly related to earlier explosions, as is clearly demonstrated by the 3D $\lnu=2.1\eft\ergs$ model, which has the highest average entropy of any model shown but does not explode within the simulated time.  In any case, since entropy depends on the integrated heating, not the heating rate, this suggests that perhaps some material is exposed longer to heating in 3D than in 2D, to which we return to in \S~\ref{sec:dwell}.

\subsection{Radial Profiles}
Moving beyond integral quantities, we can look at radial profiles, averaged over solid angle, to distinguish the global structures of 2D and 3D models.  In order to highlight the trends with luminosity and dimension, we compute two sets of profiles, one time-averaged from $200\ms$ to $300\ms$ post-bounce and another time-averaged from $450\ms$ to $550\ms$ post-bounce.  The $\lnu=2.3\eft\ergs$ models are excluded from the latter set of plots because they are already well into explosion.  Time averaging is essential for the 2D profiles to minimize the large fluctuations that obscure their underlying quasi-steady structure.

\begin{figure*}[htb]
\centering
\subfigure{\includegraphics[width=\columnwidth]{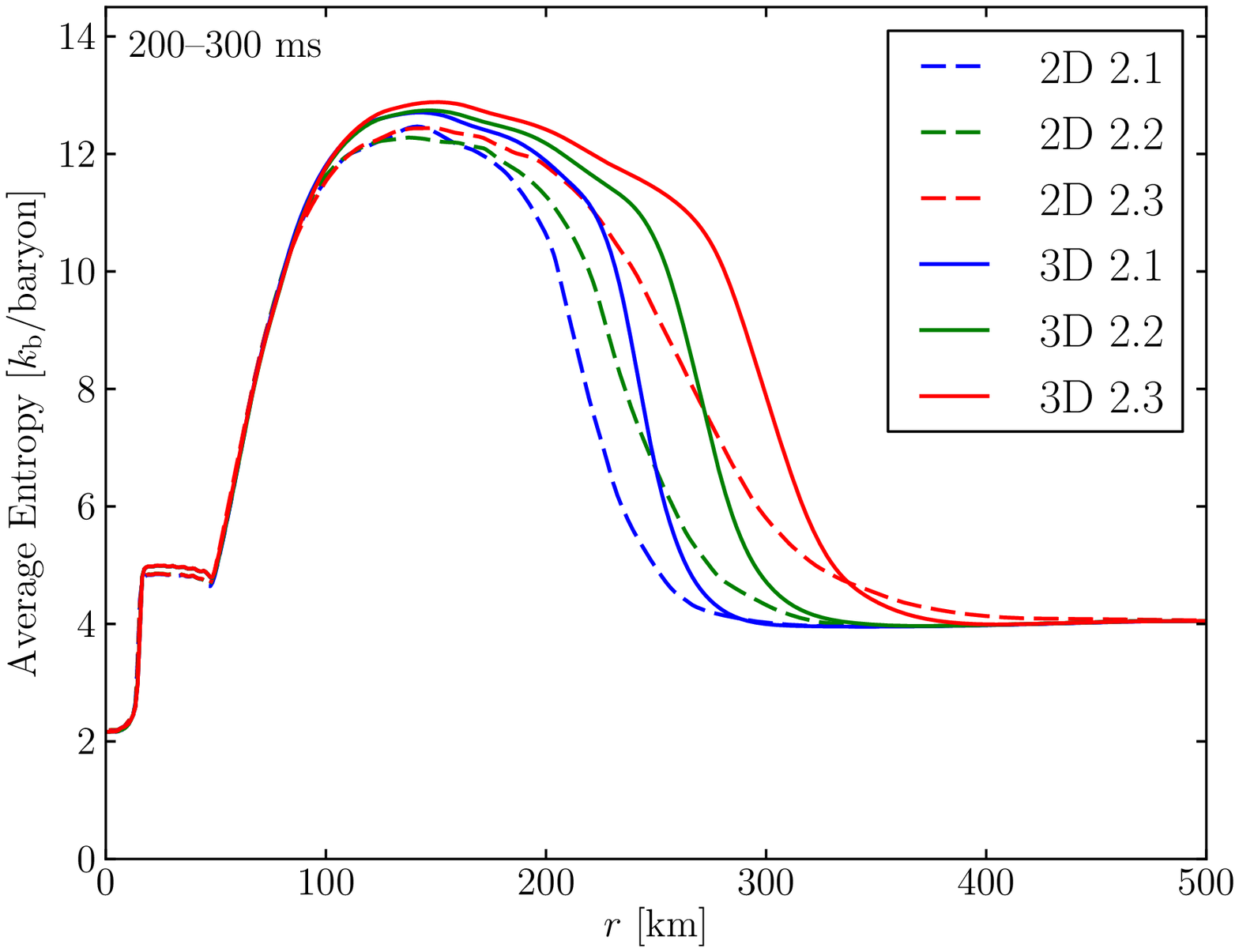}}\hfill
\subfigure{\includegraphics[width=\columnwidth]{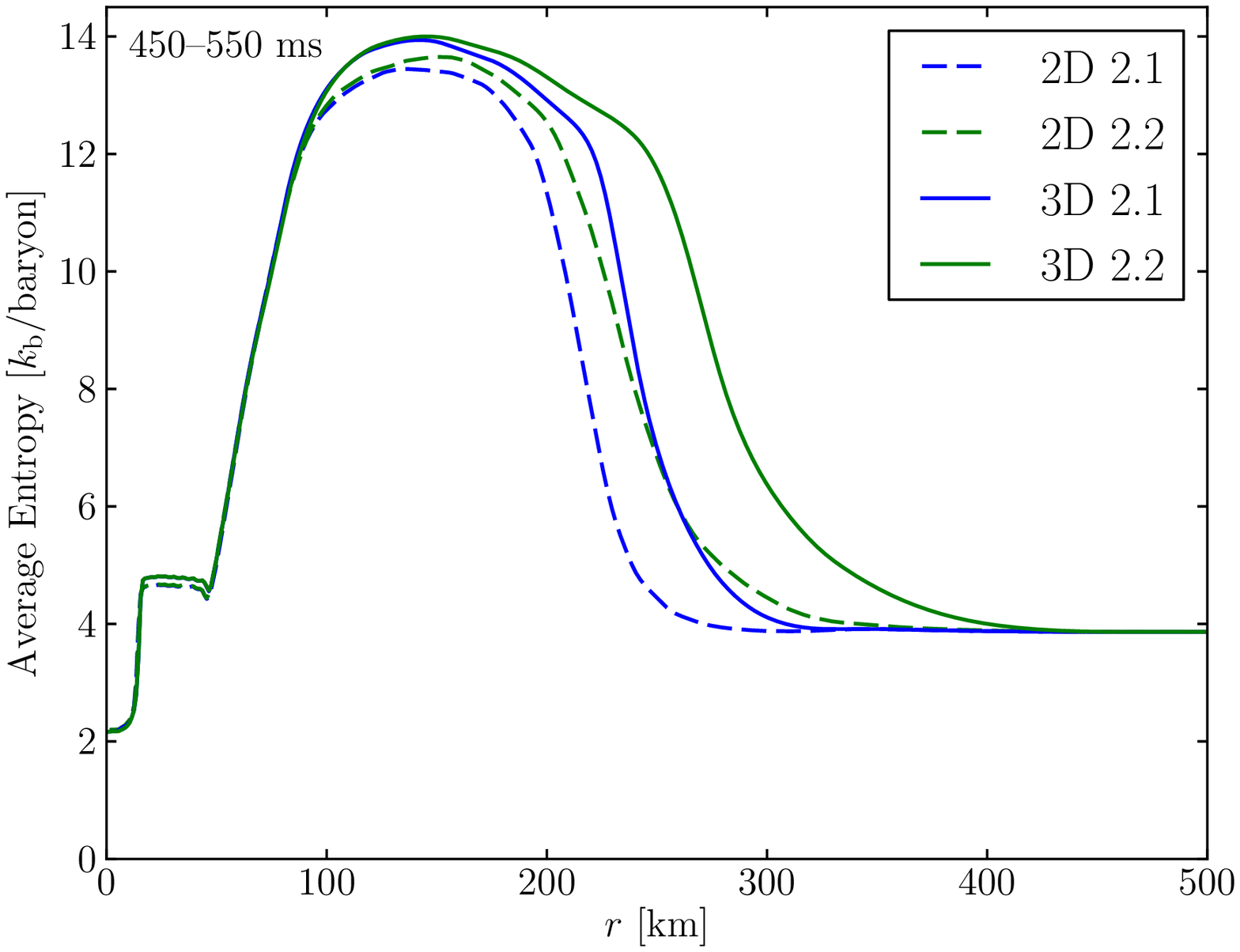}}
\caption{Time- and spherically-averaged entropy profiles in the inner $500\km$ for 2D (dashed) and 3D (solid) models.  The left panel shows the profiles time-averaged from $200$--$300\ms$ post bounce while the right panel shows the profiles time-averaged from $450$--$500\ms$ post-bounce. The $\lnu=2.3\eft\ergs$ models explode before $500\ms$ and are therefore not included in the latter.  The 3D models generically have higher averaged entropies between $\sim$$100\km$ (near the gain radius) and the shock.}
\label{fig:prof_ent}
\end{figure*}

\begin{figure*}[htb]
\centering
\subfigure{\includegraphics[width=\columnwidth]{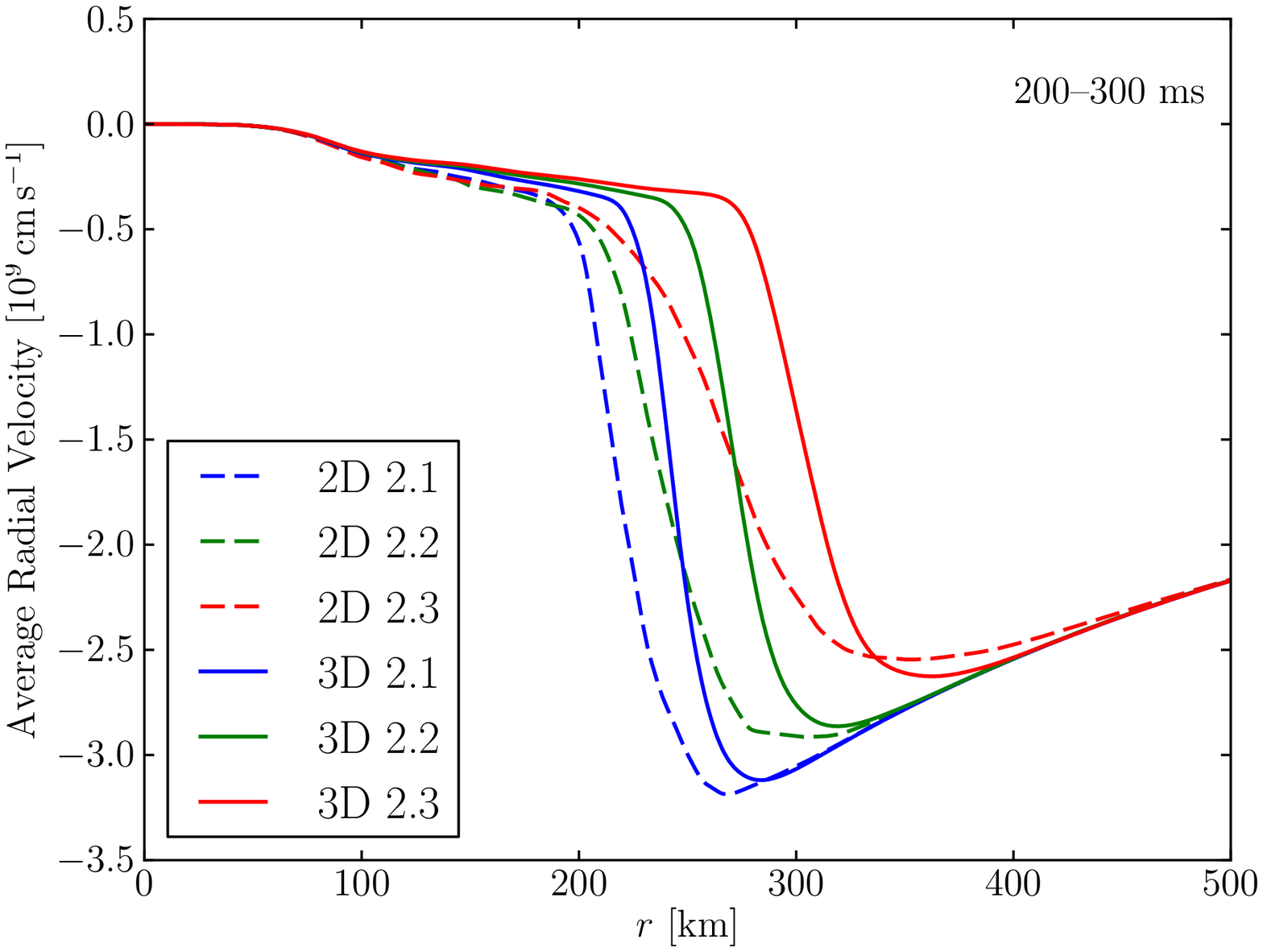}}\hfill
\subfigure{\includegraphics[width=\columnwidth]{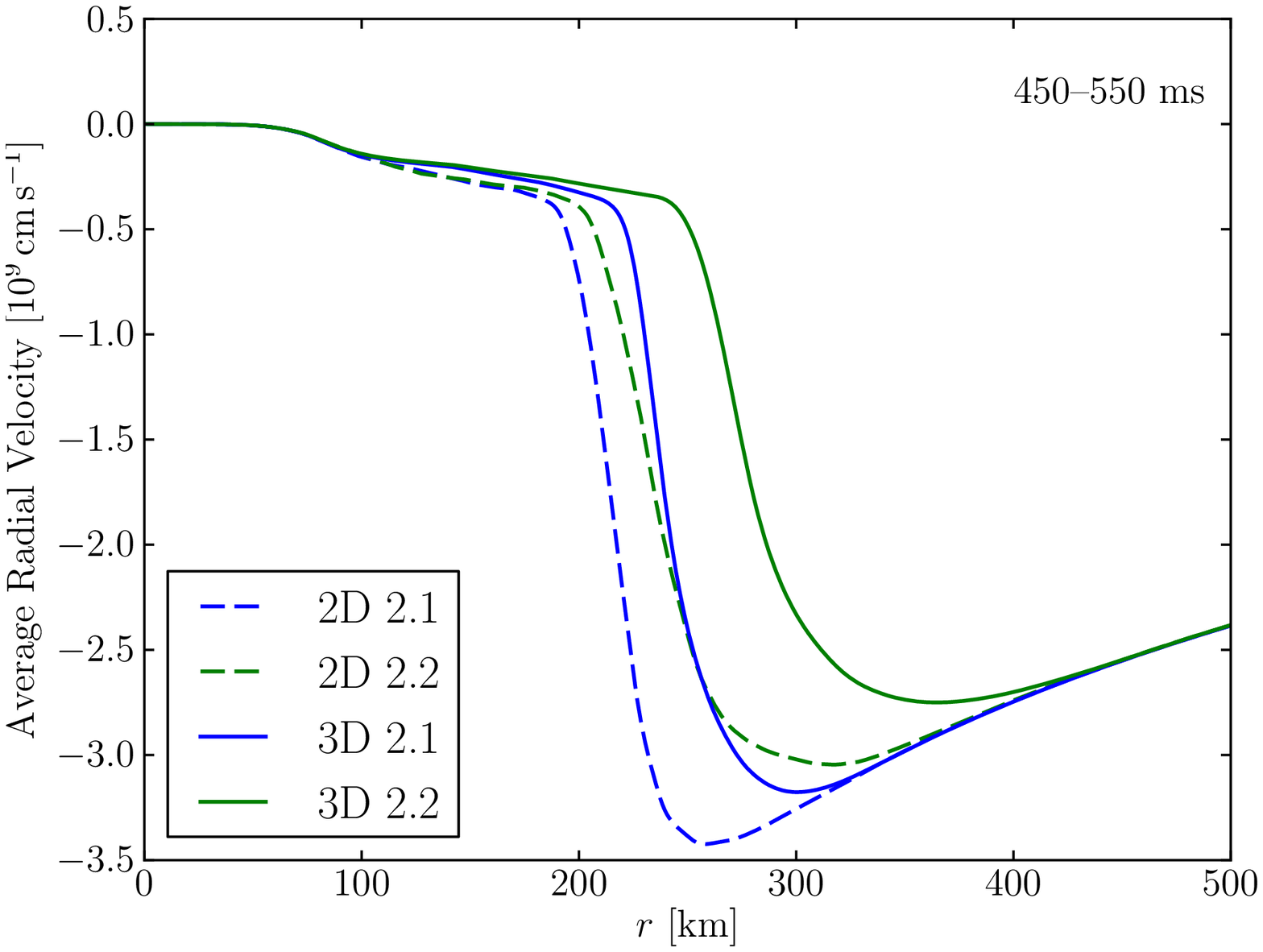}}
\caption{The same as Fig.~\ref{fig:prof_ent}, but for the time- and spherically-averaged radial velocity.  The 3D models (solid) tend to have smaller radial velocities between $\sim$$100\km$ and the shock than the 2D models (dashed).}
\label{fig:prof_vr}
\end{figure*}

\begin{figure*}[htb]
\centering
\subfigure{\includegraphics[width=\columnwidth]{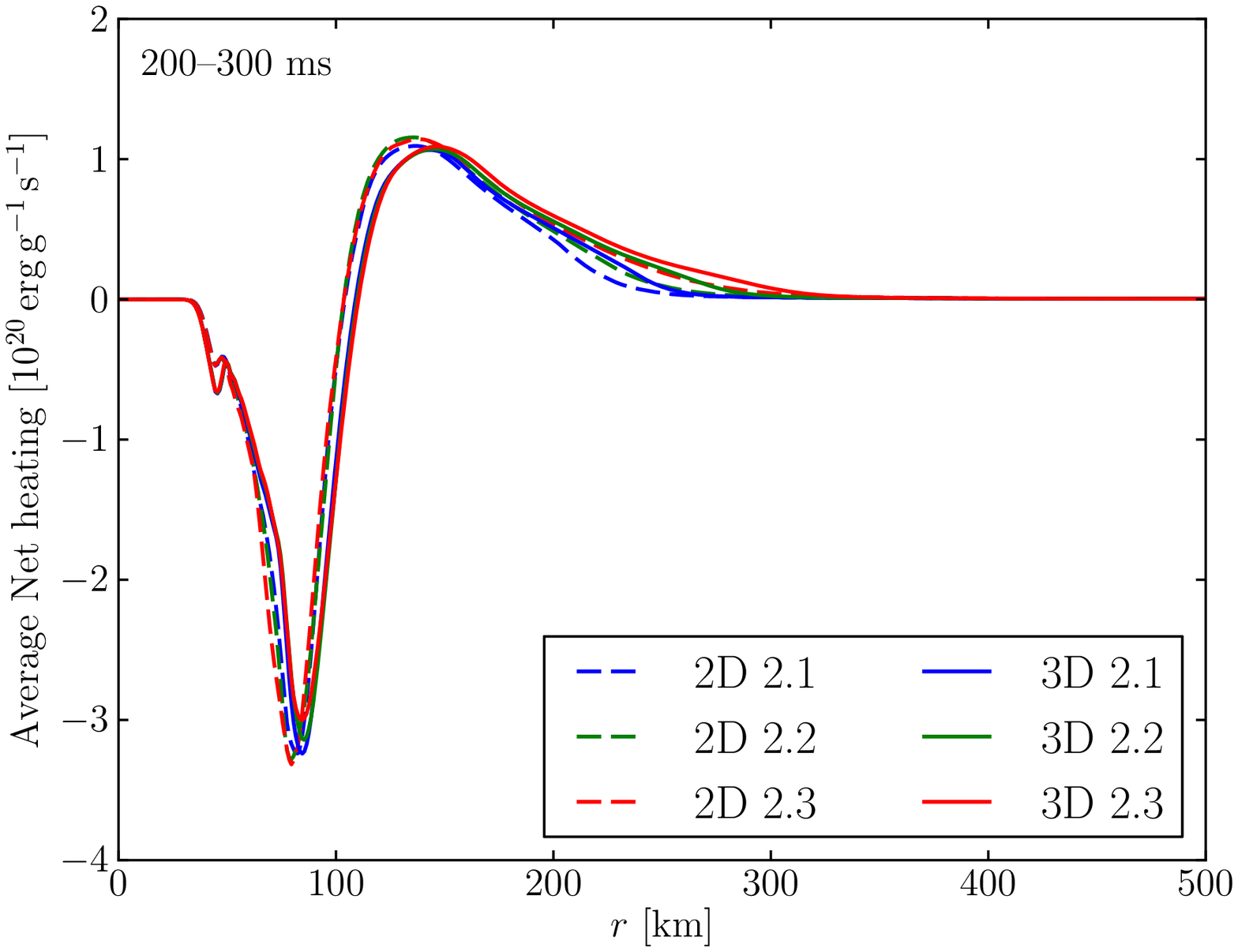}}\hfill
\subfigure{\includegraphics[width=\columnwidth]{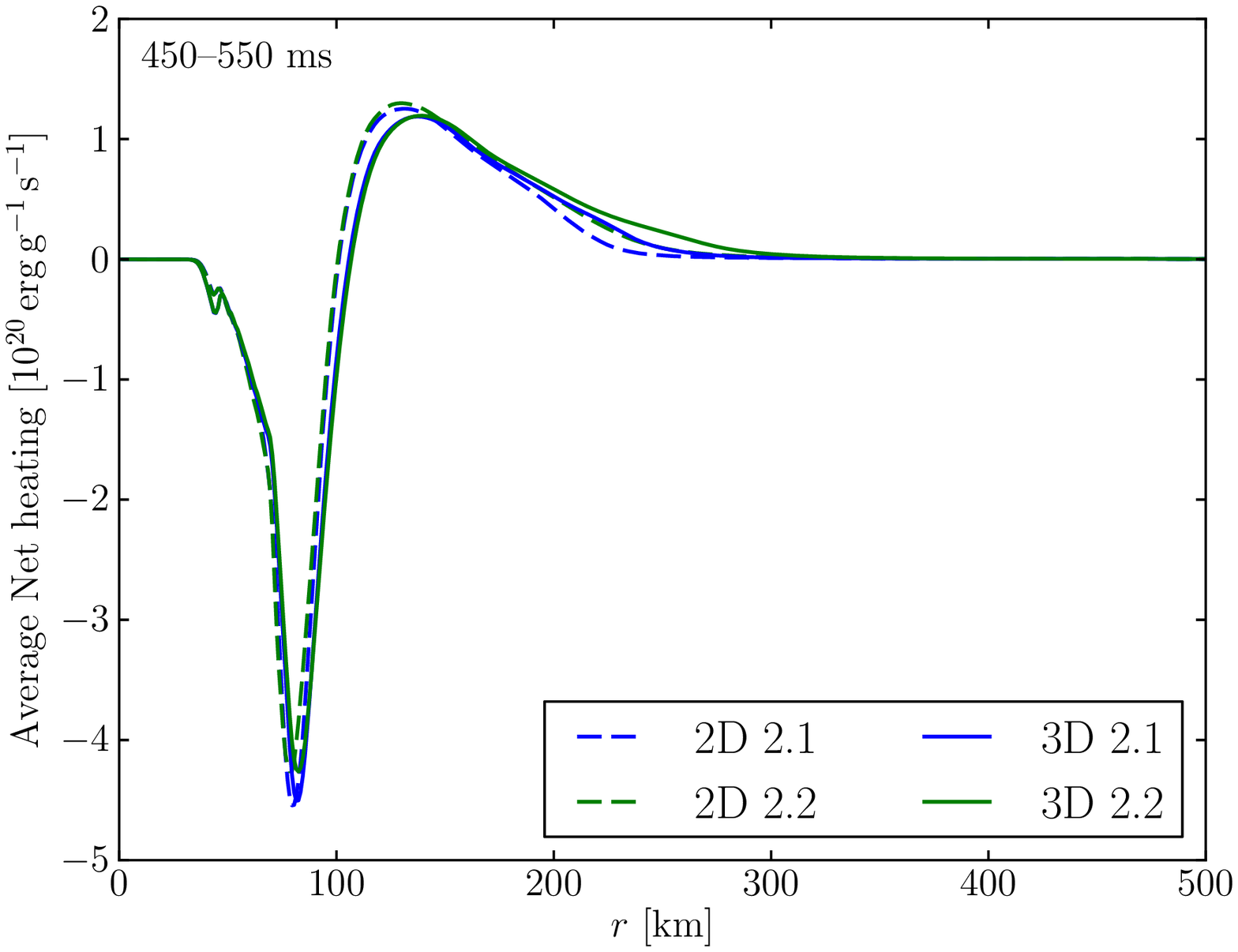}}
\caption{The same as Fig.~\ref{fig:prof_ent}, but for the time- and spherically averaged net heating rate.  Overall, the 2D (dashed) and 3D (solid) models have very similar net heating profiles.  Note that the ``gain radius'' is found at $\sim$$100\km$ where the net heating changes sign.}
\label{fig:prof_netheat}
\end{figure*}

\begin{figure*}[htb]
\centering
\subfigure{\includegraphics[width=\columnwidth]{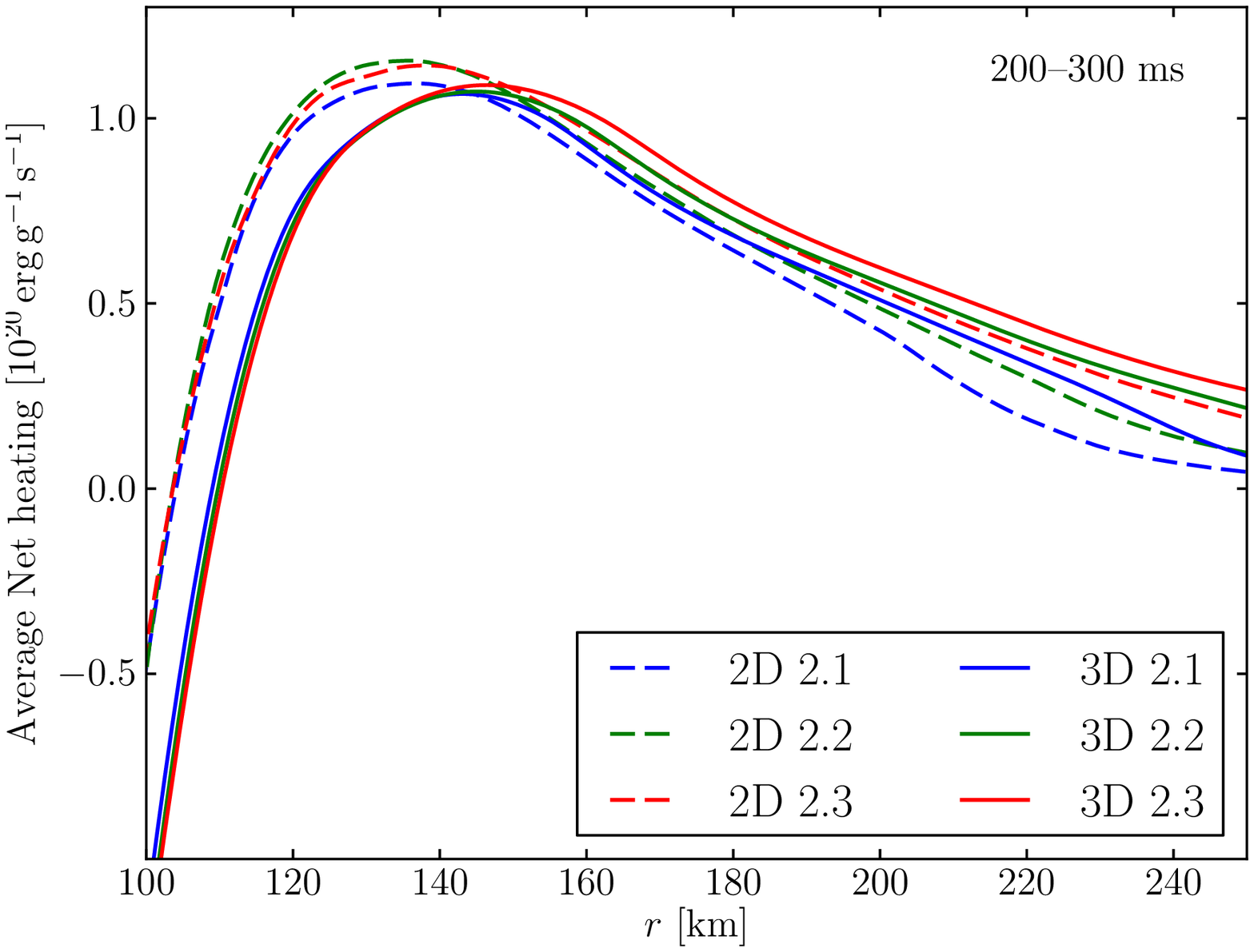}}\hfill
\subfigure{\includegraphics[width=\columnwidth]{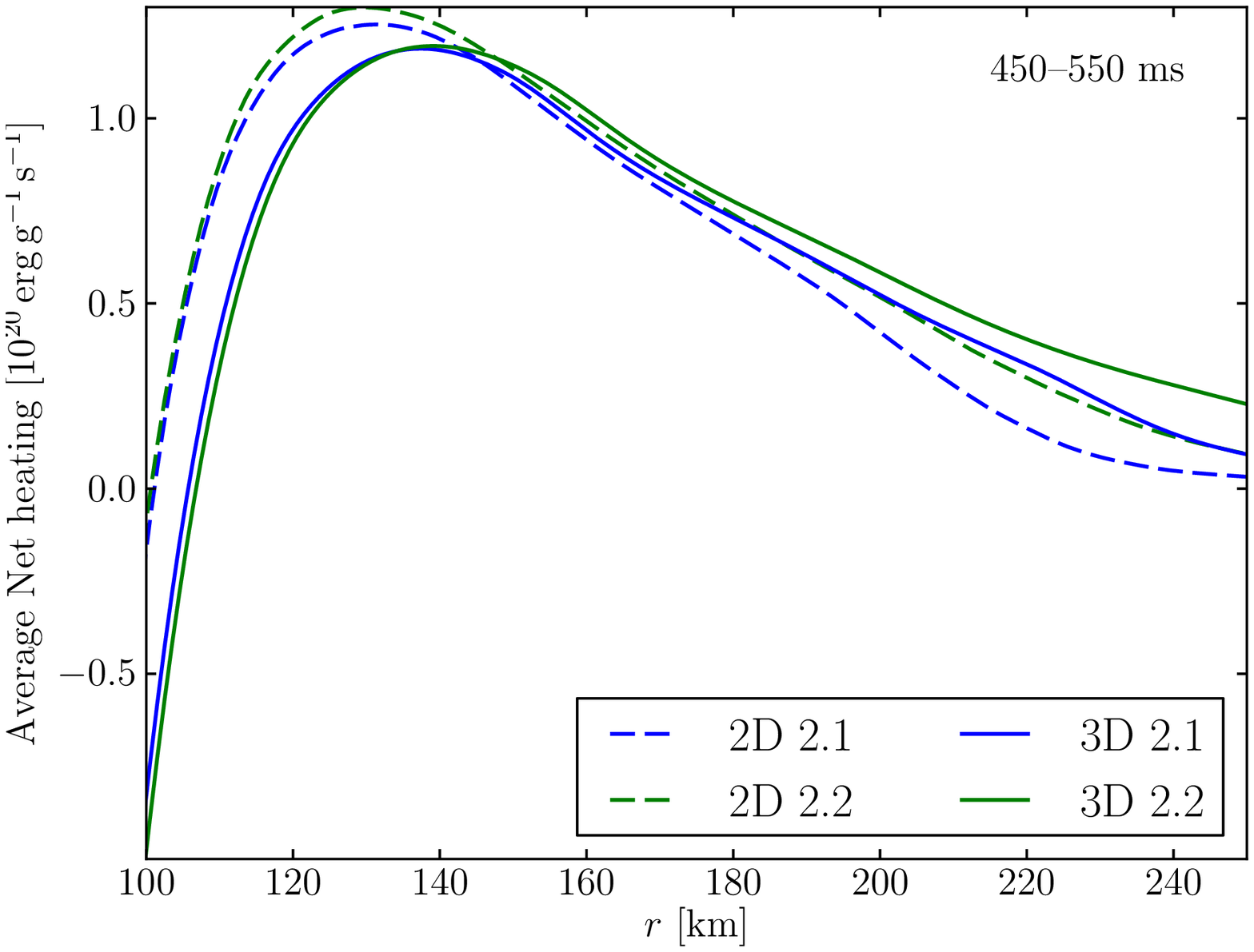}}
\caption{The same as Fig.~\ref{fig:prof_netheat}, but zoomed into the region of significant net neutrino heating.  The 2D models (dashed) tend to have more heating at smaller radii (near the gain radius where the net heating changes sign), but less heating at larger radii, compared to the 3D models (solid).}
\label{fig:prof_netheat_zoom}
\end{figure*}

\begin{figure*}[htb]
\centering
\subfigure{\includegraphics[width=\columnwidth]{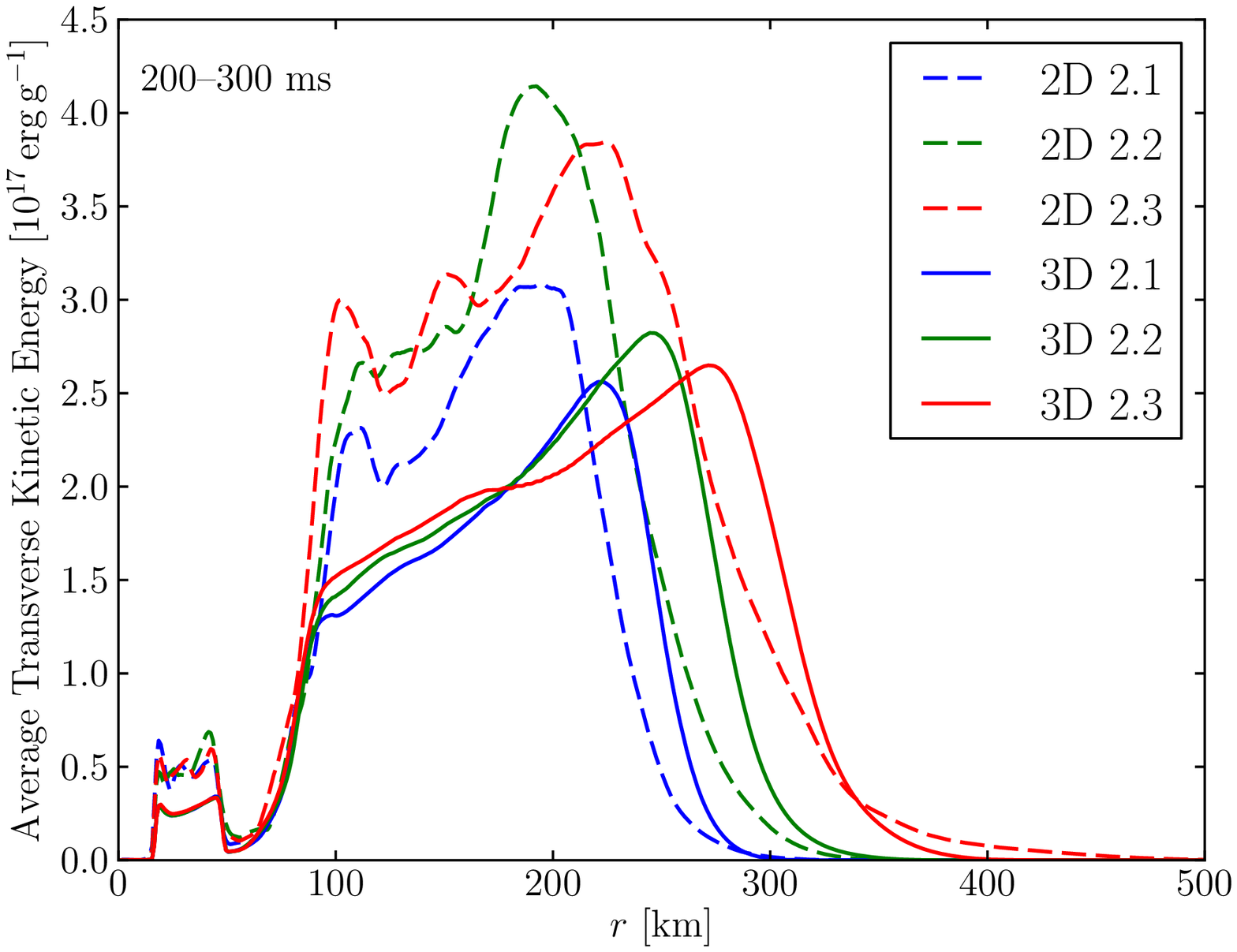}}\hfill
\subfigure{\includegraphics[width=\columnwidth]{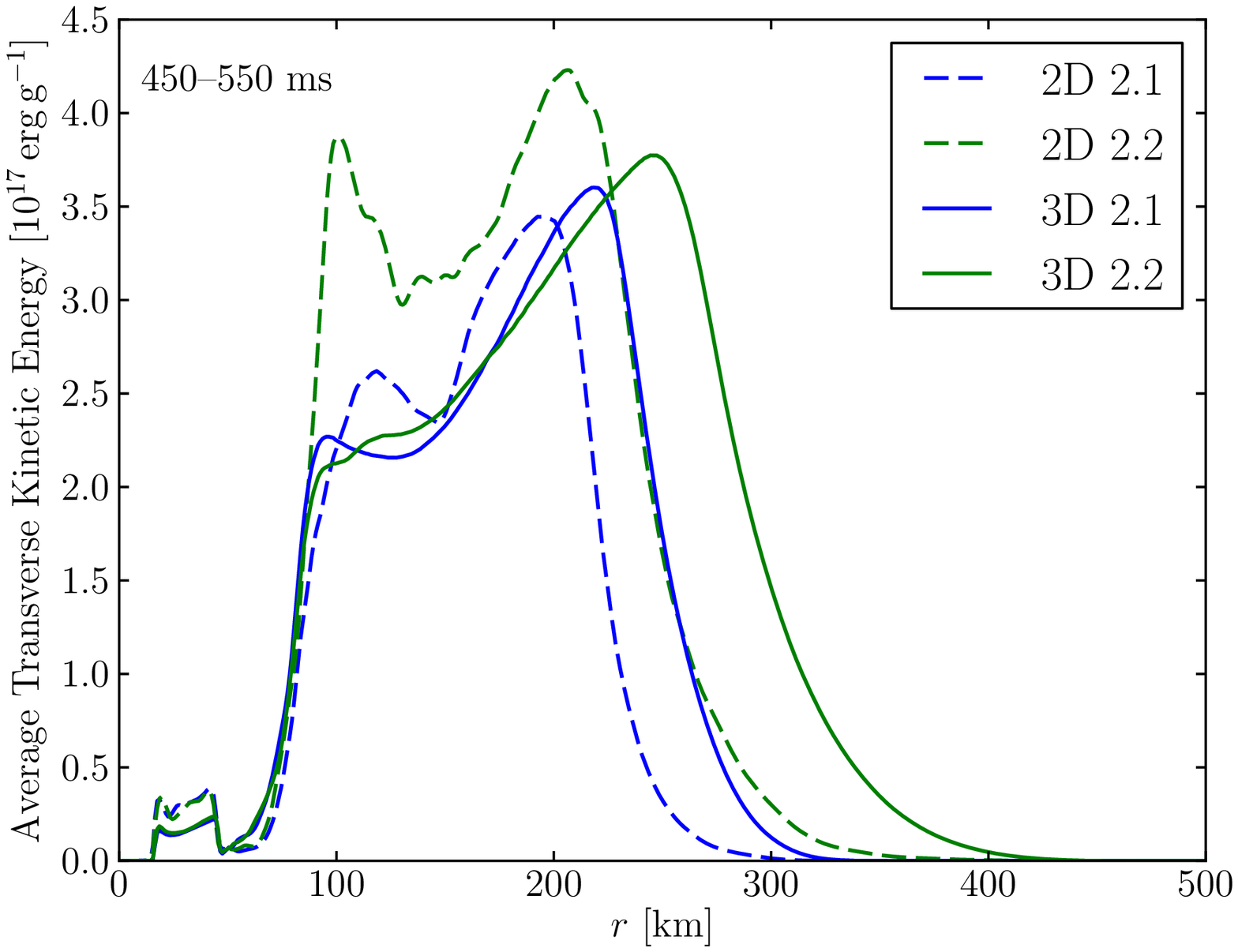}}
\caption{The same as Fig.~\ref{fig:prof_ent}, but for the time- and spherically-averaged transverse kinetic energy per unit mass.  In the early stalled accretion shock phase ($200$--$300\ms$ post-bounce), the 2D models (dashed) have $\sim$$50$--$100\%$ more specific transverse kinetic energy than the 3D models (solid).  This early turbulent vigor seems to be an \emph{artifact} of assuming axisymmetry.  At later times, the 2D/3D difference is reduced, but is still significant ($\sim$$30\%$).}
\label{fig:prof_transke}
\end{figure*}

Figure~\ref{fig:prof_ent} shows the time- and spherically-averaged entropy profiles.  Consistent with Fig.~\ref{fig:rshock}, the entropy profiles clearly show that the shock radii are systematically larger in 3D than in 2D prior to explosion.  The 3D models also have higher peak entropies than their corresponding 2D models, consistent with Fig.~\ref{fig:avgent}.  Interestingly, the peak entropies vary little over the luminosity range considered for models with a given number of dimensions, but there is a clear distinction between the 2D and 3D models.  On the other hand, the entropy profiles between $\sim$$50\km$ and $\sim$$100\km$ are remarkably similar between all models.

Figure~\ref{fig:prof_vr} shows the profiles of radial velocity.  As with the entropy profiles, the radial velocities in the post-shock region vary more with dimension than over the range of luminosities of models with a given number of dimensions.  The magnitudes of the radial velocities in the post-shock flow are generally larger in 2D than in 3D.  This remains true also when doing mass-weighted spherical averages, though the profiles are somewhat closer.  The larger post-shock radial velocities in 2D are associated with lower densities and, therefore, lower heating rates for radii beyond $\sim$$120$--$150\km$ as compared with 3D.  This has the effect of lowering the temperatures, which suppresses cooling and moves the gain radius inward (relative to 3D) towards higher densities and neutrino fluxes.  Figures~\ref{fig:prof_netheat} and \ref{fig:prof_netheat_zoom} show the radial profiles of the net neutrino heating rate and bear out these arguments.  In the end, the smaller gain radius in 2D leads to a larger integrated net heating rate in the gain region in 2D as shown in Fig.~\ref{fig:gain_net} and a larger mass in the gain region as shown in Fig.~\ref{fig:mgain}, even though the densities and net heating rates are larger in 3D at radii beyond $\sim$$120\km$ and $\sim$$150\km$, respectively.

Finally, Fig.~\ref{fig:prof_transke} shows the radial profiles of the transverse kinetic energy.  From $200\ms$ and $300\ms$ and between $\sim$$100\km$ and the shock, the 2D models have nearly twice as much transverse kinetic energy as the 3D models.  Again, the profiles vary much more with dimension than between the different luminosity models of the same dimension.  At this stage in the evolution, the turbulence in the post-shock flow is \emph{artificially} vigorous in 2D axisymmetric models, consistent with multidimensional simulations of stellar convection \citep{Meak07}.  At later times, the transverse kinetic energies in the 3D models have grown to be roughly comparable with the 2D models, though the $\lnu=2.2\eft\ergs$ models still differ by $\sim$$30\%$.

\section{Turbulence Diagnostics}\label{sec:turb}
\subsection{Dwell-Time Distributions}\label{sec:dwell}
\begin{figure}[htb]
\centering
\includegraphics[width=\columnwidth]{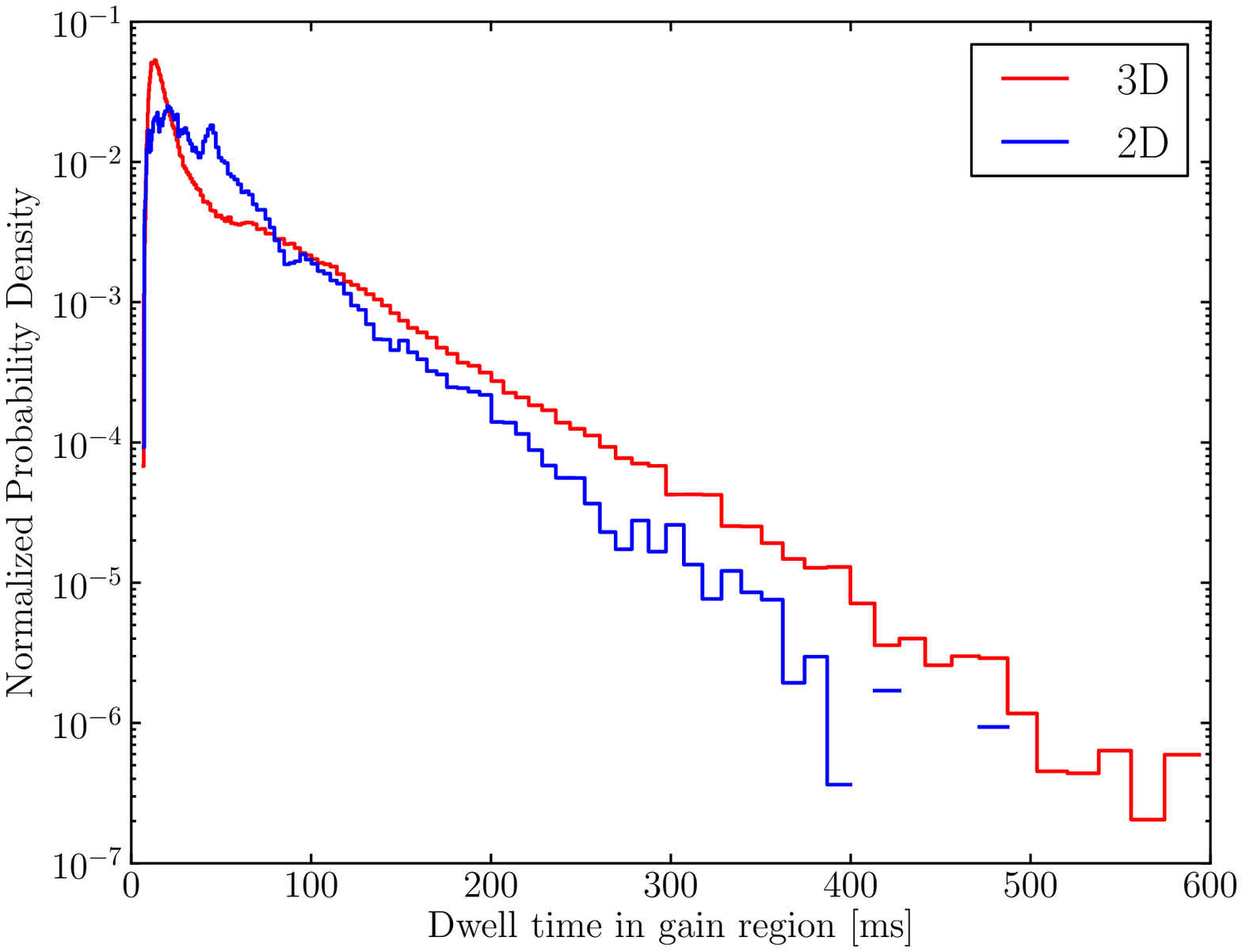}
\caption{Distribution of dwell times in the gain region for tracer particles injected at $250\ms$ post-bounce in the 2D (blue) and 3D (red) $\lnu=2.1\eft\ergs$ models.  The 2D model has a longer mean dwell time, but the 3D distribution has a long shallow tail.  Note that the data beyond $400\ms$ for the 2D model is dominated by shot noise associated with the finite number of tracer particles.}
\label{fig:dwell}
\end{figure}

\begin{figure}[htb]
\centering
\includegraphics[width=\columnwidth]{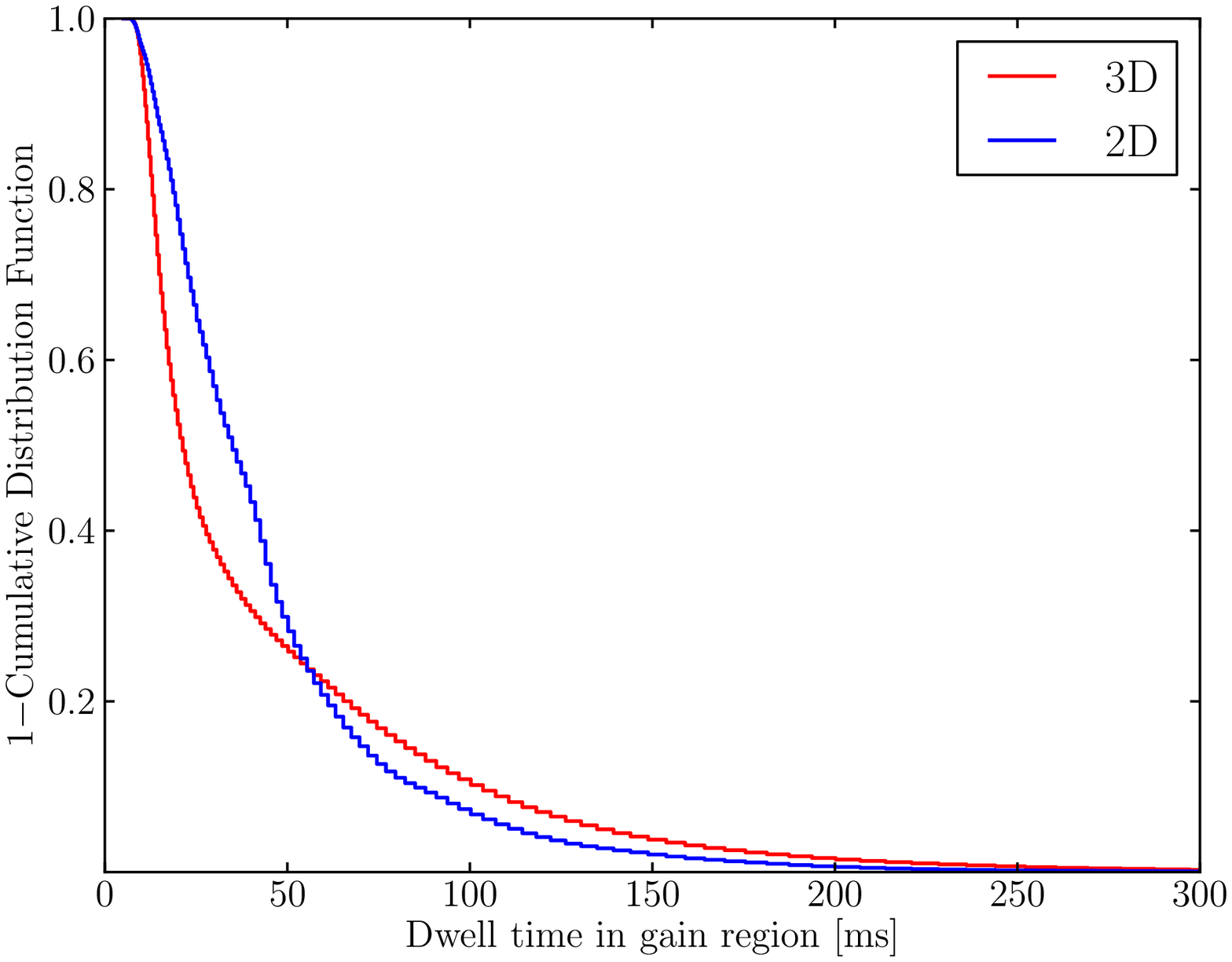}
\caption{One minus the cumulative distribution function of dwell times in the gain region for 2D (blue) and 3D (red).  The curves cross at $\sim$$50\ms$, with the 3D values larger at later times, indicating that $\sim$$25\%$ of the accreted material spends more time in the gain region in 3D compared to 2D.}
\label{fig:dwell_cum}
\end{figure}

During the stalled accretion shock phase, fluid elements advect through the gain region and eventually settle onto the proto-neutron star.  In spherical symmetry, the time to advect through the gain region (the dwell time) is short and shared by all fluid elements belonging to the same mass shell.  In multiple dimensions, the aspherical shock structure and post-shock turbulence lead to a distribution of dwell times, with some fraction of the mass being exposed longer to net neutrino heating \citep{Murp08}.  This effect can increase the neutrino heating efficiency and has naturally been suggested to be one key element of the multidimensional picture \citep{Burr95a,Murp08}.  Here, we focus on the $\lnu=2.1\eft\ergs$ models since the dwell time distribution is difficult to interpret once fluid elements begin to participate in explosion.

We measure the dwell-time distributions by following the trajectories of Lagrangian tracer particles which move according to the time-dependent velocity field.  We initialize the particles at a radius of $400\km$ at $250\ms$\footnote{We also looked at results based on particles injected at $500\ms$ post-bounce and the basic conclusions remain unchanged.} post-bounce and begin integrating their dwell times once they pass through the shock.  In 3D, we use approximately $2^{18}$ particles distributed quasi-uniformly on the sphere.  In 2D, the fewer degrees of freedom means that there are fewer independent trajectories for fluid elements of a given mass shell, even at the same resolution.  To combat this we use 16 shells of approximately $2^{12}$ particles injected over $2\ms$ for a total of $2^{16}$ particles.  In 2D, the particles are distributed uniformly in angle and we weight each particle's contribution to the dwell time distribution by $\sin\theta$ to account for the fact that each particle represents an angle-dependent annulus.

The results are shown in Fig.~\ref{fig:dwell}.  We find that there are two important effects.  First, the mean dwell time, $\langle\tau\rangle$, is longer in 2D.  In steady state, the mass in the gain region is related to the mean dwell time simply by $M_{\rm gain}=\dot{M} \langle\tau\rangle$ and so a longer mean dwell time in 2D is consistent with the larger gain mass seen previously in Fig.~\ref{fig:mgain}.  Second, the dwell time distribution in 3D has a longer, shallower tail than the 2D distribution.  Figure~\ref{fig:dwell_cum} shows one minus the cumulative distribution functions from which it can be seen that about $25\%$ of the material spends more time in the gain region in 3D than in 2D.  By recording the peak entropy reached by each tracer particle, we have found that there is a strong correlation between dwell time and peak entropy, suggesting that these long lived trajectories may be responsible, at least in part, for the larger average entropies seen in the 3D models and discussed previously.

\subsection{Turbulent Energy Spectra}
Perhaps the single most distinguishing characteristic between the 2D and 3D turbulent post-shock flows is found in their energy spectra.  As shown by \citet{Krai67} and later confirmed experimentally in various contexts \citep[see, e.g.,][]{Boff12}, turbulent cascades are different in 2D and 3D.  In 3D turbulence, energy is the only constant of the motion and this leads to a single turbulent cascade that transfers energy from some driving wavenumber $k_d$ towards larger $k$ (smaller scales).  In 2D, both energy and squared vorticity are constants of the motion which leads to cascades of energy and enstrophy (proportional to the mean squared vorticity).  The enstrophy cascade transports enstrophy in $k$-space from the driving wavenumber $k_d$ toward larger $k$ (smaller scales).  This leads to a characteristic $k^{-3}$ scaling of the velocity energy spectrum for $k > k_d$ \citep{Krai67}.  The energy cascade, by contrast, transports energy from $k_d$ to \emph{smaller} $k$ (larger scales) and leads to a $k^{-5/3}$ scaling of the velocity energy spectrum for $k < k_d$, in direct analogy with the Kolmogorov theory of turbulence.  This is the inverse energy cascade of 2D turbulence, which tends to exaggerate motions on the largest scales of the flow.  These turbulence theories were developed within highly idealized setups, assuming, for example, steady isotropic turbulence, and do not necessarily apply directly to the turbulence seen in the core-collapse context.  Nevertheless, we find, as did \citet{Hank12}, that the basic predictions of these theories---the predominance of energy at the largest scales in 2D, the excess energy at the largest scales in 2D relative to 3D, and the shallower slope of the velocity energy spectrum for $k>k_d$ (and therefore more energy for large $k$) in 3D relative to 2D---are all confirmed in our simulations.

The most natural basis to represent the matter fields in the quasi-spherical post-shock flow is the basis of real spherical harmonics.  We decompose the arbitrary scalar quantity $Q$ into spherical harmonics with time- and radially-dependent coefficients
\begin{equation}
a_{lm}(t,r) = \oint Q(t,r,\theta,\phi) Y_l^m(\theta,\phi) d\Omega,
\end{equation}
where
\begin{equation}
Y_l^m(\theta,\phi) = \begin{cases} 
	\sqrt{2} N_l^m P_l^m(\cos\theta) \cos m\phi&		m>0,\\
	N_l^0 P_l^0(\cos\theta) &				m=0,\\
	\sqrt{2} N_l^{|m|} P_l^{|m|}(\cos\theta) \sin |m|\phi&	m<0
\end{cases}
\end{equation}
and
\begin{equation}
N_l^m = \sqrt{\frac{2l+1}{4\pi}\frac{(l-m)!}{(l+m)!}}\;.
\end{equation}
In 2D, axisymmetry implies that all coefficients with $m\neq0$ are identically zero.  Here, we consider $Q=\{\rho,P,s,\sqrt{\rho}v_i\}$, where $v_i$ represents the spherical velocity components $\{v_r,v_\theta\}$ in 2D and $\{v_r,v_\theta,v_\phi\}$ in 3D.  We compute the discrete energy spectrum as a function of spherical harmonic degree $l$ as
\begin{equation}
E(l) = \sum_{m=-l}^l a_{lm}^2,
\end{equation}
where it should be understood that $a_{lm}$, and, therefore, $E(l)$, depend on time and radius \citep{Burr12}.

\begin{figure*}[htb]
\centering
\subfigure{\includegraphics[width=\columnwidth]{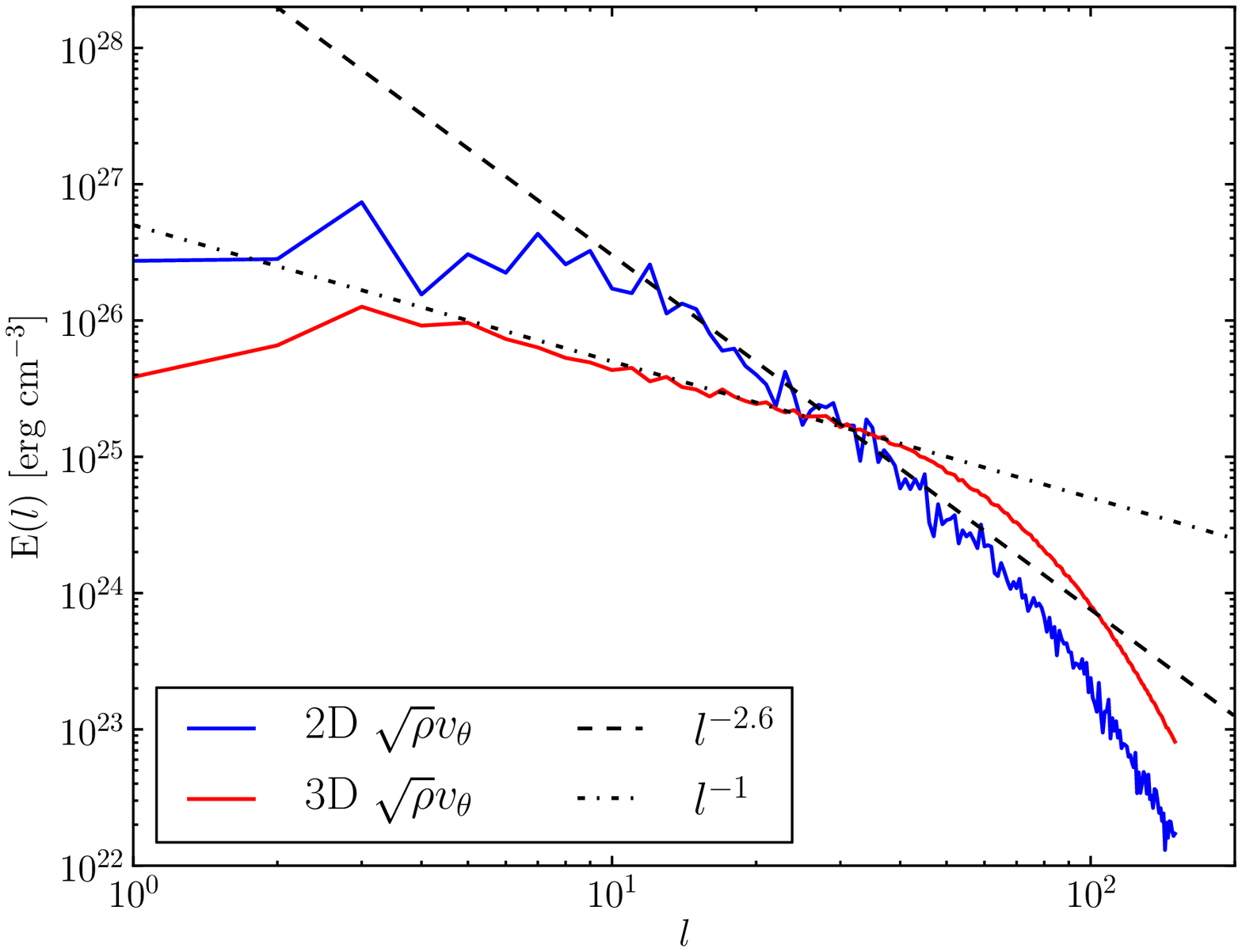}}\hfill
\subfigure{\includegraphics[width=\columnwidth]{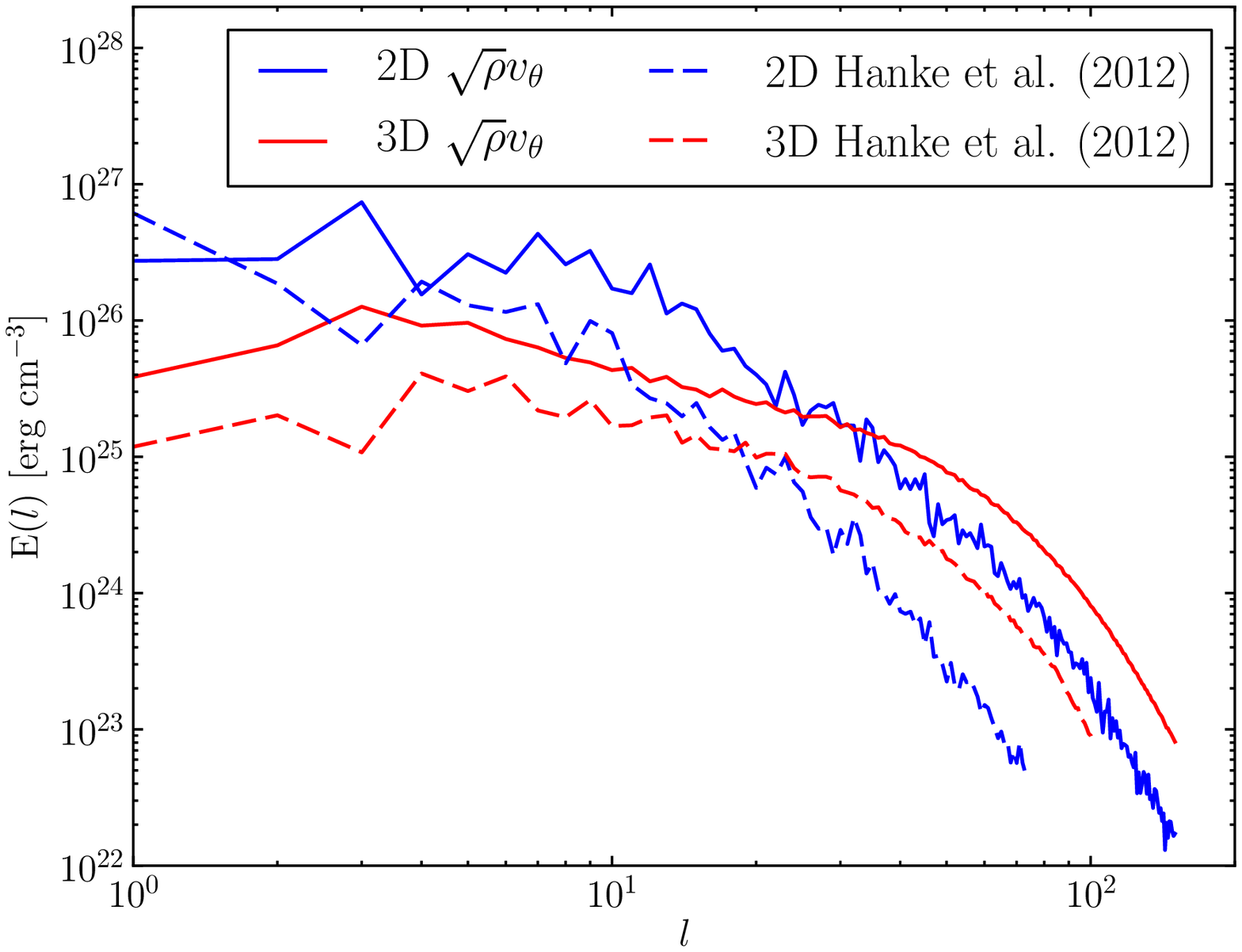}}
\caption{Discrete energy spectra of $\sqrt{\rho}v_\theta$ as a function of spherical harmonic degree $l$ for the $\lnu=2.1\eft\ergs$ 2D (blue) and 3D (red) models, measured at $r=150\km$ and time-averaged between $450$--$500\ms$ post-bounce.  The left panel includes power-laws $\propto l^{-2.6}$ and $\propto l^{-1}$, indicating the inertial range scaling in 2D and 3D, respectively.  The spectra show that the 2D model has excess power for all $l\lesssim30$.  Grid-scale dissipation sets in at $l\sim40$.  The right panel reproduces the results shown in \citet{Hank12} for comparison (dashed).  Both sets of results show the same inertial range scalings, but the excess power at low $l$ is even more extreme in the results from \citet{Hank12}.}
\label{fig:psd}
\end{figure*}

Figure~\ref{fig:psd} shows the energy spectra of $\sqrt{\rho}v_\theta$ for the $\lnu=2.1\eft\ergs$ models at a radius of $150\km$ and time-averaged between $450$--$500\ms$ after bounce.  We also reproduce the results of \citet{Hank12} for comparison.  While these curves change in detail for different quantities, over time, and at different radii, the qualitative trends and relation between 2D and 3D are quite robust.  At $l=1$, 2D has an order of magnitude more energy in all quantities $Q=\{\rho,P,s,\sqrt{\rho}v_i\}$, consistent with the idea of an inverse energy cascade which pumps energy into the largest scales.  The results of \citet{Hank12} are even more extreme, with a factor $\approx50$ more energy in the $l=1$ mode in 2D relative to 3D.  In real-space, this excess energy at the largest scales manifests as the characteristic ``sloshing'' always found in 2D simulations (including those with coupled neutrino transport).  This sloshing is typically associated with the development of the SASI \citep{Blon03,Sche08,Fogl07}, but the inverse energy cascade will always produce excess energy at $l=1$ in 2D, even if the turbulence is driven by, for example, neutrino-driven convection.  The SASI may be capable, in principle, of producing significant energy in $l=1$, but disentangling its effects from the inevitable $l=1$ energy associated with the inverse cascade would seem to require 3D simulations.  Whether the low-$l$ energy is a result of the SASI or not, all of the 3D supernova simulations thus far presented in the literature show muted or nonexistent sloshing \citep{Iwak08,Hank12,Burr12}, though there has yet to be a fully self-consistent 3D radiation hydrodynamic simulation.  Therefore, the results seen thus far suggest that these violent sloshing motions are an \emph{artifact} of assuming axisymmetry, not a feature that must be incorporated into 3D models as suggested by \citet{Hank12}.  Importantly, this artifact is not a small effect; most of the energy in the flow in 2D simulations is at low-$l$.  This artifact, in fact, dominates the flow and has poorly understood consequences on other coupled aspects of the problem, including the neutrino transport.  

At intermediate $l$, the spectra are significantly steeper in 2D ($\sim l^{-2.6}$) than in 3D ($\sim l^{-1}$).  That these power-laws differ from those naively expected from simple theoretical arguments should not be surprising given that the turbulence analyzed here is, among other things, not steady-state or isotropic.  Qualitatively, however, the expectation of a steeper slope in 2D relative to 3D is confirmed and we find our results generally consistent with those of \citet{Hank12}\footnote{\citet{Hank12} report scalings of $l^{-3}$ and $l^{-5/3}$ in 2D and 3D, respectively, but Fig.~\ref{fig:psd} shows that their results are consistent with the shallower slopes reported here.}.  The transition to these slopes occurs around $l\sim10$ in 2D and $l\sim4$ in 3D and reflects a characteristic scale for convective plumes, a scale that likely depends on the model-dependent size of the gain region.  The 3D model has significantly more energy at small-scales than the 2D model.  At $l\gtrsim40$ (spatial scales $\sim$$10\km$ at a radius of $150\km$), we begin to see the effects of grid-scale ($2\km$ at this radius) dissipation.

\citet{Hank12} showed that their results, especially in 3D, were sensitive to resolution.  The comparison of energy spectra in Fig.~\ref{fig:psd} affords us the opportunity to directly compare the effective resolutions of our independent calculations by comparing the scales at which dissipation begins to set in.  As noted above, our 3D results begin to deviate from the $l^{-1}$ power-law at $l\approx40$, while the results in \citet{Hank12} begin to deviate at $l\approx25$, confirming our expectation that our models have nearly double the effective resolution.  \citet{Hank12} argue that 3D models become less prone to explosion as resolution is increased, yet our higher resolution 3D models explode earlier than the 2D models.  The source of this apparent discrepancy is unclear.

\section{Multidimensional Explosions}\label{sec:expl}
\subsection{Explosion Conditions}
\begin{figure}[htb]
\centering
\includegraphics[width=\columnwidth]{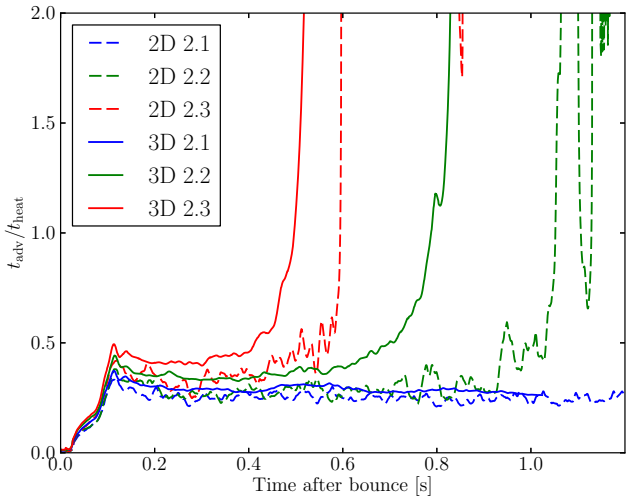}
\caption{Ratio of advection to heating timescales (defined in the text) for the 3D (solid) and 2D (dashed) models.  Explosions are associated with a rapid growth in this ratio, though it is difficult to identify a particular critical value for the onset of explosions.  With a critical value of $0.5$, the $\lnu=2.3\eft\ergs$ model explodes $159\ms$ earlier in 3D relative to 2D, while the $\lnu=2.2\eft\ergs$ model explodes $333\ms$ earlier in 3D.}
\label{fig:adv_heat_ratio}
\end{figure}

A number of quantities have been proposed in the literature that are meant to distinguish exploding from non-exploding models and, further, to define in a systematic way the time at which a model transitions into the exploding phase.  The most widely used condition is based on the idea of a critical ratio of the advection to heating time scales \citep{Jank98,Thom00a,Thom03}.  There are numerous ways of defining these timescales.  Here, we define the advection time as
\begin{equation}
t_{\rm adv} = \int_{R_{s}}^{R_{\rm gain}} \frac{dr}{\langle v_r \rangle}
\end{equation}
where $\langle v_r\rangle$ is the spherically-averaged radial velocity.  In multi-D models, the shock radius $R_s$ and the gain radius $R_{\rm gain}$ are not uniquely defined bounds\footnote{Indeed, the gain region need not even be bounded from below by a single closed surface.}.  We appeal to the radial profiles shown previously and define the shock radius as the outermost zero in the radial velocity gradient and the gain radius as the first zero crossing in the net heating rate interior to the shock (always around $\sim$$100\km$).  Furthermore, to minimize the large fluctuations in $t_{\rm adv}$ that appear in 2D due to transient and localized fluctuations in $v_r$, we time-average the velocity profiles over a $\pm15\ms$ window.  We define the heating timescale as
\begin{equation}
t_{\rm heat} = \frac{\int_{R_{\rm gain}}^{R_{s}} (\langle\rho\varepsilon\rangle - \langle\rho\rangle\varepsilon_0(\langle\rho\rangle,\langle Y_e\rangle)) 4 \pi r^2 dr}{\int_{R_{\rm gain}}^{R_{s}} \langle\rho\rangle (\langle\mathcal{H}-\mathcal{C}\rangle) 4 \pi r^2 dr},
\end{equation}
where angle brackets indicate solid-angle averaging, $\varepsilon$ is the specific internal energy, $\varepsilon_0$ is the zero-point energy of the EOS, and $\mathcal{H}-\mathcal{C}$ is the net heating rate per unit mass.  The results are shown in Fig.~\ref{fig:adv_heat_ratio}.  After an initial transient phase, the models settle on an approximately (model-dependent) constant value.  Rather than there being a particular critical value for the ratio of advection to heating time scales, explosions seem to be robustly connected to rapid growth from these model-dependent quasi-steady values.  Nonetheless, we define an explosion time in a systematic manner by measuring the time at which this ratio exceeds $0.5$ without later dropping below this value.

\begin{figure}[htb]
\centering
\includegraphics[width=\columnwidth]{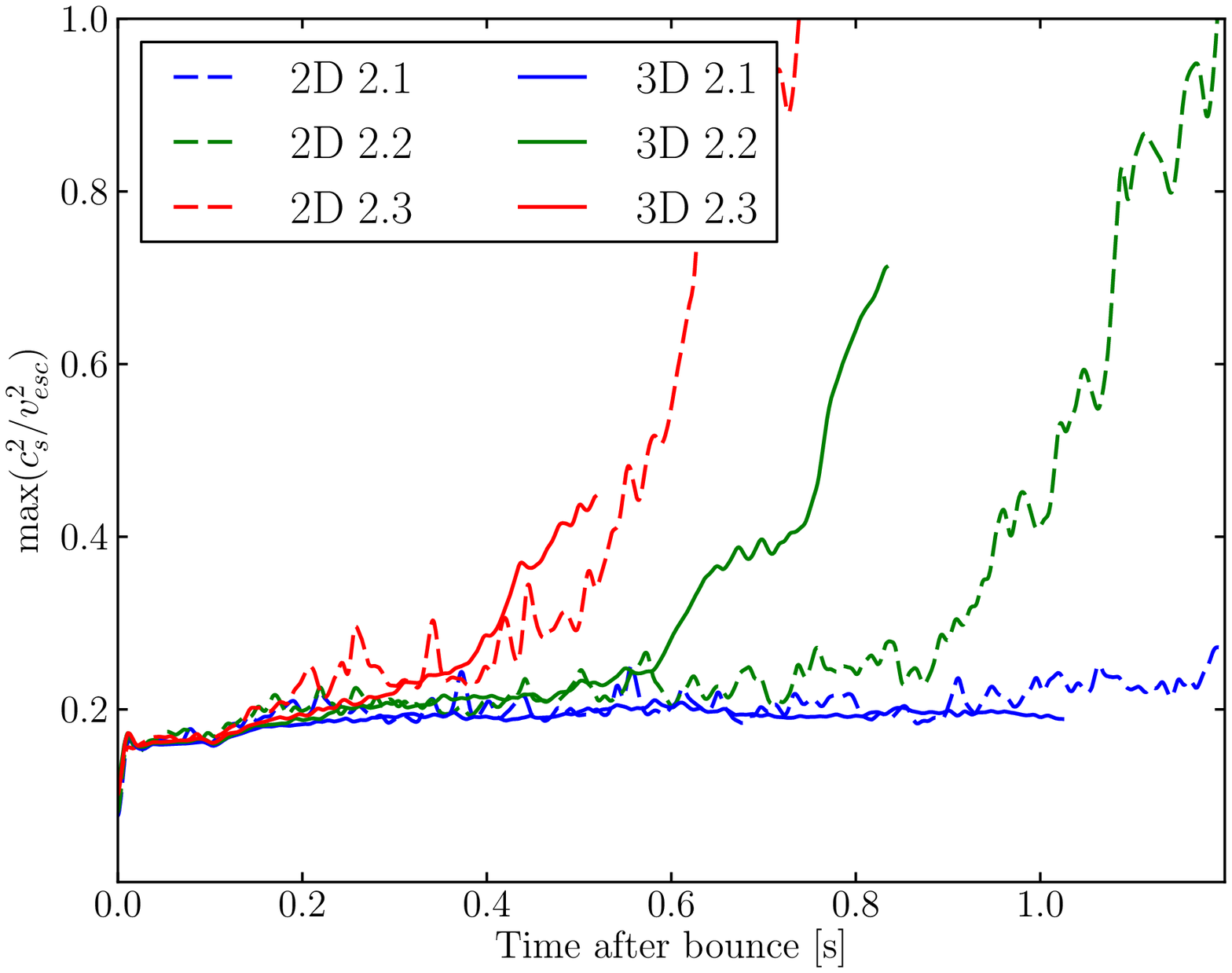}
\caption{Maximum of the ratio of the square of the local sound speed to escape speed for the 3D (solid) and 2D (dashed) models, smoothed for clarity, as discussed by \citet{Pejc12} in their antesonic condition.  In agreement with \citet{Mull12}, we find that the critical value of $0.2$ suggested by \citet{Pejc12} is too low, but using a larger value ($\approx0.3$) may be a useful diagnostic of explosion.   With a critical value of $0.3$, the $\lnu=2.3\eft\ergs$ model explodes $87\ms$ earlier in 3D relative to 2D, while the $\lnu=2.2\eft\ergs$ model explodes $309\ms$ earlier in 3D.}
\label{fig:ante}
\end{figure}

An alternative explosion condition (the ``antesonic'' condition) was suggested by \cite{Pejc12}.  Based on parametrized 1D steady-state models, they suggested that explosions occur when ${\rm max}(c_s^2/v_{\rm esc}^2)$ reaches a critical value $\approx 0.2$.  \citet{Murp11} tested this condition with their parametrized 2D simulations and found it to be consistent with their results.  \citet{Mull12} on the other hand, using 2D radiation hydrodynamic simulations, assessed the validity of the antesonic condition and suggest that it may not be a robust indicator of explosion, but, in any case, the critical value should at least be somewhat larger ($\sim$$0.35$).  Like \citet{Mull12}, we find that a larger value for the critical condition is required and we adopt $0.3$ as the critical ratio.  Unfortunately, ${\rm max}(c_s^2/v_{\rm esc}^2)$ is a noisy quantity, with brief spikes as large as $\approx0.5$ that then return well below the critical value.  The smoothed curves, however, seem to reliably indicate when explosions set in, but these smoothed curves are necessarily produced \textit{ex post facto}.  In other words, the antesonic condition is not a reliable indicator of explosion given the instantaneous value of the ratio.  The evolutions for the 2D and 3D models, smoothed for clarity, are shown in Fig.~\ref{fig:ante}.  As above, we identify the time of explosion as the time when the smoothed version of ${\rm max}(c_s^2/v_{\rm esc}^2)$ exceeds $0.3$ without later dropping below this value.

A final, phenomenological, explosion condition is simply when the average shock radius exceeds $400\km$ without later receding below this value, as used in \citet{Nord10}.  In this case, explosion times can be read directly from the shock radius evolution curves in Fig.~\ref{fig:rshock}.

\begin{deluxetable*}{lccccccc}
\tabletypesize{\footnotesize}
\tablewidth{0pt}
\tablecaption{Explosion Times and Accretion Rates\label{tab:exp}}
\tablecolumns{8}
\tablehead{
\colhead{$L_{\nu_e}$\tablenotemark{a}} & \colhead{Dimension} & \multicolumn{6}{c}{Explosion Condition}\\
\cline{1-8}\\[-1.5ex]
\colhead{} & \colhead{} &\multicolumn{2}{c}{$t_{\rm adv}/t_{\rm heat}$} & \multicolumn{2}{c}{${\rm max}(c_s^2/v_{\rm esc}^2)$} & \multicolumn{2}{c}{$\langle R_s\rangle$}\\[0.3ex]
\cline{1-8}\\[-2ex]
\colhead{} & \colhead{} & \colhead{$t_{\rm exp}$} & \colhead{$\dot{M}$} & \colhead{$t_{\rm exp}$} & \colhead{$\dot{M}$} & \colhead{$t_{\rm exp}$} & \colhead{$\dot{M}$}\\
\colhead{} & \colhead{} & \colhead{(ms)} & \colhead{($\msun\,{\rm s}^{-1}$)} & \colhead{(ms)} & \colhead{($\msun\,{\rm s}^{-1}$)} & \colhead{(ms)} & \colhead{($\msun\,{\rm s}^{-1}$)}
}
\startdata
\multirow{2}{*}{2.2} & 2D & 1031 & 0.182 & 920 & 0.196 & 954 & 0.188\\
& 3D & 698 & 0.213 & 611 & 0.224 & 695 & 0.214\\[1em]
\multirow{2}{*}{2.3} & 2D & 571 & 0.231 & 501 & 0.241 & 557 & 0.233\\
& 3D & 412 & 0.269 & 414 & 0.268 & 434 & 0.262
\enddata
\tablenotetext{a}{$10^{52}\ergs$}
\end{deluxetable*}

Table~\ref{tab:exp} shows the explosion times and corresponding mass accretion rates as determined by the three conditions above for the $\lnu=2.2\eft\ergs$ and $\lnu=2.3\eft\ergs$ models.  All three conditions give comparable numbers, but the antesonic condition tends to give the earliest indication of explosion.  We note, however, that this conclusion may be somewhat sensitive to the particular critical values adopted.  That the three conditions give comparable explosion times is not surprising; the ratio of advection to heating times, the ratio of sound to escape speed, and the average shock radius all grow as models transition into explosion.  None of these conditions is able to robustly predict when or even if an explosion will occur before the explosion begins.  This suggests that while these conditions may be indicative of explosion, they are by no means the full story.

With the results shown in Table~\ref{tab:exp} in hand, we can quantify the delay between the 2D explosions and the earlier 3D explosions.  Taking the minimum difference found between the explosion conditions, we find that the $\lnu=2.3\eft\ergs$ 3D model explodes at least $87\ms$ before the corresponding 2D model, while the $\lnu=2.2\eft\ergs$ 3D model explodes at least $259\ms$ earlier than the corresponding 2D model.  These delays may translate into appreciable differences in the explosion energy at infinity \citep{Yama12}, with earlier explosions (3D) being more energetic.

\subsection{Explosion Trigger in 3D}
What conditions trigger explosions in 3D?  The multi-D nature of the hydrodynamics is generally agreed to be important in producing such conditions.  Turbulence in the post-shock flow leads to a distribution of dwell times for accreting parcels of matter, increasing the mass in the gain region and, therefore, the efficiency of neutrino heating.  In this way, turbulence plays a crucial, but secondary, role as an aid to neutrino heating.  \citet{Murp11} argue that convection modifies the quasi-steady global structure of the flow by introducing turbulent fluxes of, for example, enthalpy and entropy.  Their model focuses on how turbulent convection modifies averaged radial profiles, rather than on the effects of particular turbulent fluctuations, and was able to account for the differences in, for example, the radial profiles of entropy between 1D and 2D models.  While these are important roles for the post-shock turbulent motions, here we suggest that the turbulent fluctuations themselves, driven by neutrino heating, may be instrumental in triggering explosions.

\begin{figure*}[htb]
\centering
\includegraphics[width=\textwidth]{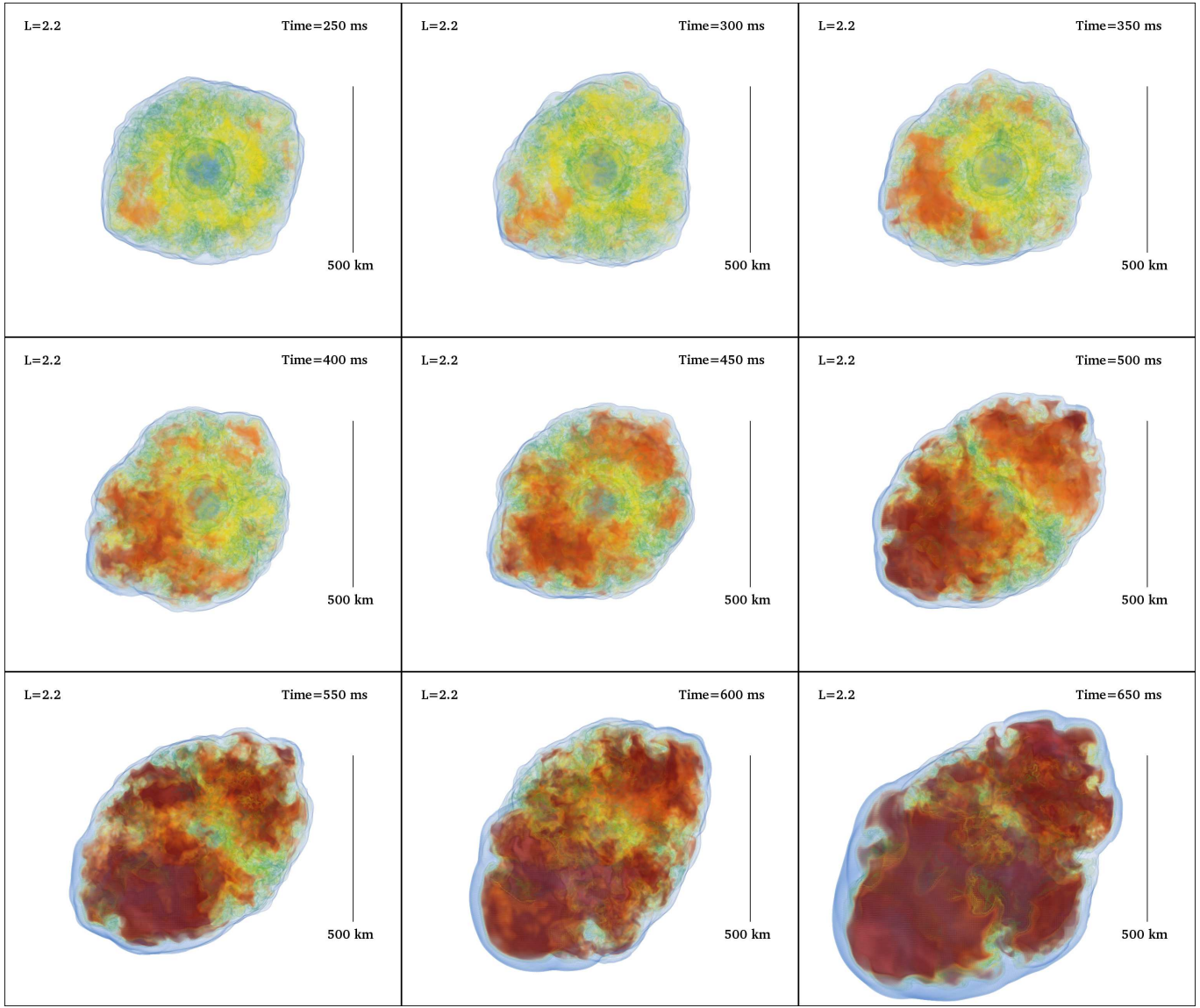}
\caption{Sequence of volume renderings of the specific entropy for the 3D $\lnu=2.2\eft\ergs$ model.  As seen at the bottom left of each image, a high entropy plume forms around $\sim$$250$--$300\ms$ and persists for hundreds of milliseconds, eventually pushing the shock out far enough to seemingly trigger the global explosion.  A similar structure appears around $\sim$$450\ms$ at the top right of each image, which leads to similar local shock expansion thereafter.}
\label{fig:volseq}
\end{figure*}

\begin{figure*}[htb]
\centering
\includegraphics[width=\textwidth]{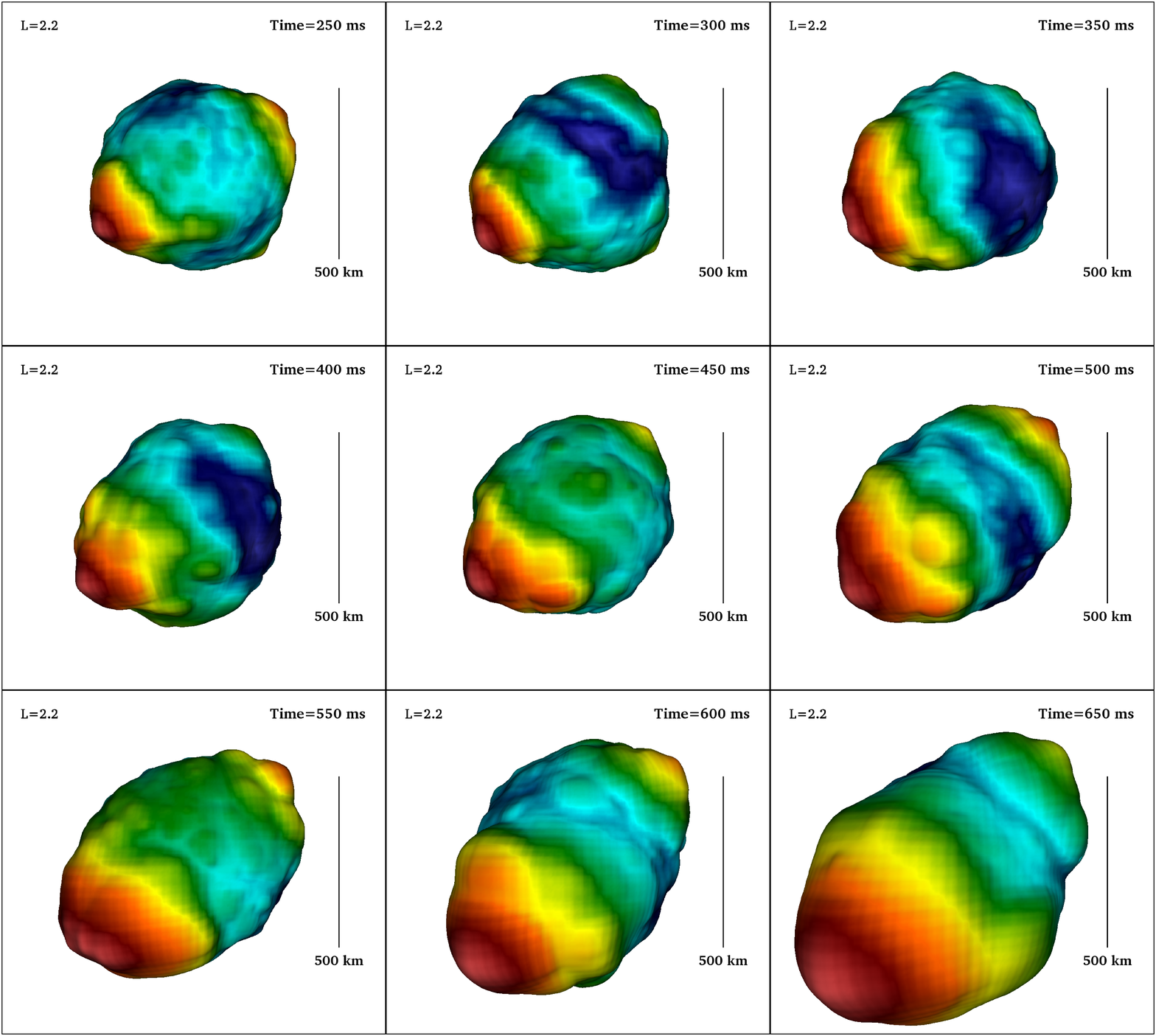}
\caption{Evolution of the shock surface from the 3D $\lnu=2.2\eft\ergs$ model.  The colors indicate radius to emphasize the aspherical nature of the shock surface and the secularly growing feature associated with the high-entropy plume shown in Fig.~\ref{fig:volseq}.}
\label{fig:shocksurf}
\end{figure*}

The dominant hydrodynamic instability, aside from the explosion itself, in our parametrized 3D models is neutrino-heating-driven convection \citep{Burr12,Murp12}.  In the nonlinear phase, the post-shock turbulent flow involves a complex interaction between buoyantly rising, neutrino-heating-driven plumes, negatively buoyant accretion streams, turbulent entrainment resulting from secondary Kelvin-Helmholtz instabilities, and dissipative interactions between rising plumes and the bounding shock.  The evolution of a buoyant ``bubble'' depends on parameters, including the heating rate, bubble size, and background inflow velocity.  Small bubbles tend to be shredded by turbulent entrainment, while large bubbles can rise all the way to the shock and locally push it outward.  Our simulations suggest that some bubbles can continue rising, locally pushing the shock to ever larger radii until the global explosion is triggered.  Figure~\ref{fig:volseq} shows a sequence of volume renderings illustrating that large buoyant features form in the flow and push the shock to larger radii.  Figure~\ref{fig:shocksurf} shows the evolution of the shock surface where the same local growth of the shock radius can be clearly seen.  Interestingly, the growth of these features occurs not on a dynamical time, but appears to proceed quasi-statically.  The same basic picture holds in all of our 3D simulations, suggesting that this may be a generic feature of the transition to explosion.

We can try to understand the conditions for the runaway growth of bubbles (and the associated triggered explosions) by considering the highly simplified toy model presented in the Appendix.  In this model, the evolution of a bubble's outer radius ($R_b$) is determined by the competition between neutrino driving power ($L_\nu \tau$, $L_\nu$ is the neutrino driving luminosity and $\tau$ is the effective optical depth of the bubble) and accretion power ($\alpha G M \dot{M}/R_b$, $\alpha$ is a constant defined in the Appendix) associated with the ram pressure of material immediately behind the shock.  The evolution follows
\begin{equation}
\frac{dR_b}{dt} = \frac{\Omega_0}{4 \pi} \frac{R_b^2}{G M M_b} \left(L_\nu \tau - \alpha \frac{G M \dot{M}}{R_b}\right),
\end{equation}
which has the solution
\begin{equation}\label{eq:rbub}
R_b(t) = \frac{\alpha G M \dot{M}}{L_\nu \tau + e^{\lambda t} (\alpha GM\dot{M}/R_0 - L_\nu \tau)},
\end{equation}
where 
\begin{equation}
\lambda = \alpha\frac{\Omega_0}{4\pi} \frac{\dot{M}}{M_b}.
\end{equation}
Here, $\Omega_0$ is the bubble's constant solid angle and $M_b$ is the bubble's fixed mass.  Qualitatively, $R_b(t)$ has two types of behavior.  When the neutrino driving term exceeds the ram pressure term, $R_b(t)$ grows without bound, presumably triggering the global explosion \citep{Thom00a}.  When the ram pressure term dominates, the bubble recedes, perhaps allowing another bubble to take its place.

\begin{figure}[htb]
\centering
\includegraphics[width=\columnwidth]{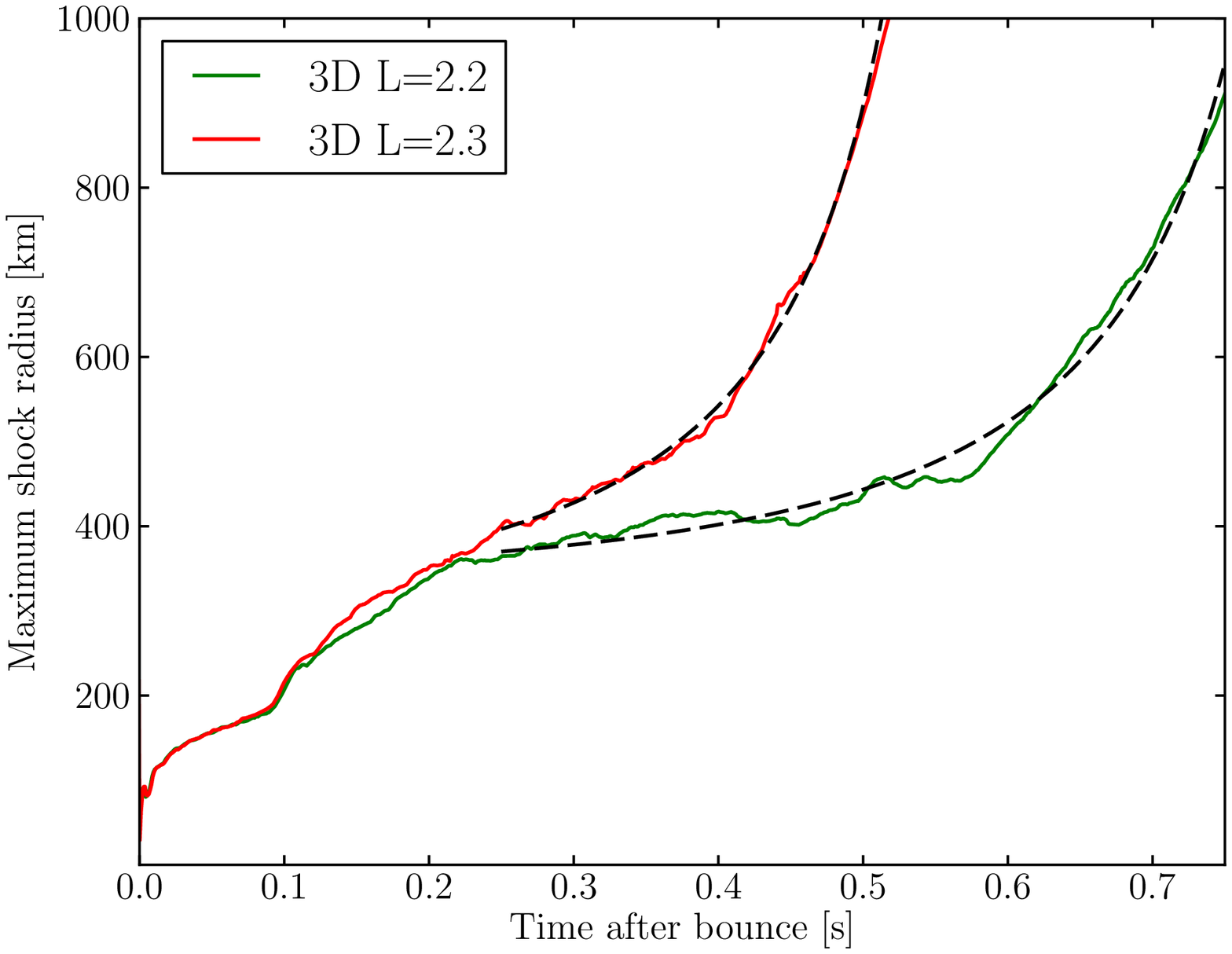}
\caption{Maximum shock radii for the 3D $\lnu=2.2\eft\ergs$ and $\lnu=2.3\eft\ergs$ models, along with fitted solutions (black dashed lines) based on Eq.~\ref{eq:rbub}.  See the text and Appendix for discussions.}
\label{fig:rmax_model}
\end{figure}

This simple model makes two predictions.  First, the model predicts a characteristic growth of the maximum shock radius, at least in the quasi-static growth phase.  Figure~\ref{fig:rmax_model} shows the maximum shock radii ($R_{\rm max}$) from the $\lnu=2.2\eft\ergs$ and $\lnu=2.3\eft\ergs$ 3D models along with fitted model solutions.  To compute the fits, we fix $M=1.6\msun$, $\dot{M}=0.25\msun\,{\rm s}^{-1}$, and $M_b/\Omega_0=2.5\times10^{30}\,{\rm g}$ and fit the model solution to the data between $250\ms$ post-bounce and the time when $R_{\rm max}=1000\km$, leaving $R_0$, $\tau$, and $\alpha$ as free parameters.  Both fits give $\tau\approx0.04$ and $\alpha\approx0.25$.  Given the simplicity of the model, the agreement between the hydrodynamical simulation results and the model predictions is quite surprising and encouraging.  

Second, there is a critical luminosity for a given mass accretion rate and shock radius above which a bubble will runaway.  If the stalled shock radius scales as $L^\beta$, then the critical luminosity for runaway bubble growth is proportional to $\dot{M}^{1/1+\beta}$.  Empirically, $\beta\sim3$ and if we adopt the parameters used above we find
\begin{equation}
\begin{split}
L_{\rm crit} \approx& 2.2 \left(\frac{M}{1.6\msun}\right)^{1/4} \left(\frac{\dot{M}}{0.25\msun\,{\rm s}^{-1}}\right)^{1/4}\\
 & \times \left(\frac{\tau}{0.04}\right)^{-1/4}\times10^{52}\ergs,
\end{split}
\end{equation}
which seems consistent with the critical explosion curves from parametrized multi-D models shown in other works \citep{Murp08,Nord10,Hank12,Couc12}.  Inasmuch as runway bubble growth is associated with explosions, this may be viewed as an alternative, albeit crude, derivation of the critical luminosity curve of \citet{Burr93a}.  This may suggest that the reduction in the critical luminosity in going from 1D to 2D and 3D models might arise, in part, from the emergence of bubbles and their runaway growth.

\section{Discussion and Conclusions}\label{sec:conc}
We have presented analyses of parametrized 2D axisymmetric and 3D core-collapse supernova models.  Our basic conclusion is not surprising nor controversial---the hydrodynamics of core-collapse supernovae are different between 2D axisymmetric and 3D models.  These differences are many and, while some are subtle and perhaps not crucial to understanding the mechanism, some are quite dramatic and make interpreting 2D supernova models problematic.

Our parametrized models indicate that the global structures of the flows are different between 2D and 3D.  We see this reflected in integral measures like the mass and average entropy in the gain region and in the radial profiles of various quantities.  In spite of identical heating and cooling prescriptions between 2D and 3D, our analyses show how the different global structures effect the heating and cooling rates.  We find that the 2D models have significantly higher integrated net heating rates than their corresponding 3D models and attribute this to the smaller gain radius (and more mass in the gain region) in 2D, which may in turn be related to the larger radial velocities in 2D.  On the other hand, the 3D models tend to have higher densities at large radii and larger average shock radii.

The dwell time distributions for accreted parcels of matter offer another vantage point from which to distinguish 2D and 3D models.  In both cases, the post-shock turbulence leads to a broad distribution of dwell times.  In our simulations, the mean dwell time is somewhat longer in 2D than in 3D, consistent with the larger mass in the gain region in 2D than in 3D, but the 3D distribution has a relatively prominent tail towards long dwell times.  About $25\%$ of the material spends more time in the gain region in 3D than in 2D, being exposed to more integrated heating and reaching higher peak entropies.

Ultimately, many of the differences we see are plausibly associated with the character of the post-shock turbulent flow.  In 2D, turbulent energy is pumped into the largest scales of the flow, which inevitably gives rise to the sloshing behavior manifest in all modern 2D core-collapse supernova simulations.  This sloshing behavior has been identified as a crucial ingredient in nearly all successful neutrino-driven explosion models to date \citep{Bura06,Mare09,Mull12}.  In 3D, this sloshing motion is muted or absent, at least in part, because the turbulent energy transport is predominantly towards small scales.  Since we see explosions earlier in 3D than in 2D, vigorous sloshing is either not critical in any dimension or the explosion mechanism operates differently in 2D and 3D.

Finally, we present a toy model that describes the evolution of buoyant bubbles driven by the competition between neutrino heating power and accretion power.  The simple model is able to account for the quasi-static growth of the maximum shock radius in the exploding 3D models with neutrino luminosities of $\lnu=2.2\eft\ergs$ and $\lnu=2.3\eft\ergs$.  It also predicts the existence of a critical luminosity for a given mass accretion rate and shock radius beyond which a bubble will have runaway growth.  We speculate that this runaway growth triggers the global explosion of the star, and therefore that the critical luminosities for runaway bubble growth and explosion are effectively the same.

The mean background and fluctuating turbulent components of the post-shock flow in 2D axisymmetric models are different from 3D models and, more importantly, likely not representative of the hydrodynamics of supernova cores in Nature.  While the models presented in this work are incomplete, lacking adequate neutrino transport and feedback, this basic conclusion seems robust.  It seems inevitable that we must await 3D models before drawing conclusions concerning the trigger of core-collapse supernovae.

\acknowledgements
The authors acknowledge stimulating interactions with Christian Ott, Rodrigo Fernandez, Thierry Foglizzo, John Blondin, Ann Almgren, and John Bell.  A.B. acknowledges support from the Scientific Discovery through Advanced Computing (SciDAC) program of the DOE, under grant number DE-FG02-08ER41544, the NSF under the sub- award No. ND201387 to the Joint Institute for Nuclear Astrophysics (JINA, NSF PHY-0822648), and the NSF PetaApps program, under award OCI-0905046 via a subaward No. 44592 from Louisiana State University to Princeton University. The authors thank the members of the Center for Computational Sciences and Engineering (CCSE) at LBNL for their invaluable support for CASTRO. The authors employed computational resources provided by the TIGRESS high performance computer center at Princeton University, which is jointly supported by the Princeton Institute for Computational Science and Engineering (PICSciE) and the Princeton University Office of Information Technology; by the National Energy Research Scientific Computing Center (NERSC), which is supported by the Office
of Science of the US Department of Energy under contract DE-AC03-76SF00098; and on the Kraken supercomputer, hosted at NICS and provided by the National Science Foundation through the TeraGrid Advanced Support Program under grant number TG-AST100001.

\appendix

\section{Toy Model of a Neutrino-Driven Bubble}\label{sec:appendix}
Consider a bubble of fixed solid angle $\Omega_0$ pinned beneath the shock at radius $R_b$.  The energy injection rate due to neutrino heating is $\sim$$(\Omega_0/4\pi) L_\nu \tau$.  For a quasi-steady system, this energy injection rate should be balanced by the rate of energy loss.  The interaction between the bubble and the surrounding flow will remove energy from the bubble at a rate $\sim C_d \rho v^3 \Omega_0 R_b^2$, where $C_d$ is an effective drag coefficient that encapsulates information about the geometry and character of the flow, including effects like turbulent entrainment.  For a bubble pinned beneath the shock, the density and velocity are given by the shock jump conditions, so this loss term can be rewritten as $\sim C_d (\Omega_0/4\pi) (f^2/\mathcal{R}^2) (2 G M \dot{M}/R_b)$, where $f$ is the pre-shock fraction of the free-fall velocity ($\approx 1/\sqrt{2}$) and $\mathcal{R}$ is the shock compression ratio.  Also, lifting the bubble out of the gravitational potential requires a power $\sim dR_b/dt GMM_b/R_b^2$.  Equating the driving neutrino power to the sum of the ram and gravitational powers, we find
\begin{equation}
\frac{dR_b}{dt} = \frac{\Omega_0}{4 \pi} \frac{R_b^2}{G M M_b} \left(L_\nu \tau - \alpha \frac{G M \dot{M}}{R_b}\right)
\end{equation}
where $\alpha=2 C_d f^2/\mathcal{R}^2$ is a dimensionless constant of order unity.

\vspace{1em}
\bibliographystyle{apj}
\bibliography{refs}

\begin{thebibliography}{38}
\expandafter\ifx\csname natexlab\endcsname\relax\def\natexlab#1{#1}\fi

\bibitem[{{Almgren} {et~al.}(2010){Almgren}, {Beckner}, {Bell}, {Day},
  {Howell}, {Joggerst}, {Lijewski}, {Nonaka}, {Singer}, \& {Zingale}}]{Almg10}
{Almgren}, A.~S., {et~al.} 2010, \apj, 715, 1221

\bibitem[{{Bethe}(1990)}]{Beth90}
{Bethe}, H.~A. 1990, Reviews of Modern Physics, 62, 801

\bibitem[{{Blondin} {et~al.}(2003){Blondin}, {Mezzacappa}, \&
  {DeMarino}}]{Blon03}
{Blondin}, J.~M., {Mezzacappa}, A., \& {DeMarino}, C. 2003, \apj, 584, 971

\bibitem[{{Boffetta} \& {Ecke}(2012)}]{Boff12}
{Boffetta}, G., \& {Ecke}, R.~E. 2012, Annual Review of Fluid Mechanics, 44,
  427

\bibitem[{{Bruenn} {et~al.}(2009){Bruenn}, {Mezzacappa}, {Hix}, {Blondin},
  {Marronetti}, {Messer}, {Dirk}, \& {Yoshida}}]{Brue09}
{Bruenn}, S.~W., {Mezzacappa}, A., {Hix}, W.~R., {Blondin}, J.~M.,
  {Marronetti}, P., {Messer}, O.~E.~B., {Dirk}, C.~J., \& {Yoshida}, S. 2009,
  Journal of Physics Conference Series, 180, 012018

\bibitem[{{Buras} {et~al.}(2006){Buras}, {Janka}, {Rampp}, \&
  {Kifonidis}}]{Bura06}
{Buras}, R., {Janka}, H.-T., {Rampp}, M., \& {Kifonidis}, K. 2006, \aap, 457,
  281

\bibitem[{{Burrows} {et~al.}(2012){Burrows}, {Dolence}, \& {Murphy}}]{Burr12}
{Burrows}, A., {Dolence}, J.~C., \& {Murphy}, J.~W. 2012, \apj, 759, 5

\bibitem[{{Burrows} \& {Fryxell}(1993)}]{Burr93}
{Burrows}, A., \& {Fryxell}, B.~A. 1993, \apjl, 418, L33

\bibitem[{{Burrows} \& {Goshy}(1993)}]{Burr93a}
{Burrows}, A., \& {Goshy}, J. 1993, \apjl, 416, L75

\bibitem[{{Burrows} {et~al.}(1995){Burrows}, {Hayes}, \& {Fryxell}}]{Burr95a}
{Burrows}, A., {Hayes}, J., \& {Fryxell}, B.~A. 1995, \apj, 450, 830

\bibitem[{{Couch}(2012)}]{Couc12}
{Couch}, S.~M. 2012, ArXiv e-prints

\bibitem[{{Foglizzo} {et~al.}(2007){Foglizzo}, {Galletti}, {Scheck}, \&
  {Janka}}]{Fogl07}
{Foglizzo}, T., {Galletti}, P., {Scheck}, L., \& {Janka}, H.-T. 2007, \apj,
  654, 1006

\bibitem[{{Hanke} {et~al.}(2012){Hanke}, {Marek}, {M{\"u}ller}, \&
  {Janka}}]{Hank12}
{Hanke}, F., {Marek}, A., {M{\"u}ller}, B., \& {Janka}, H.-T. 2012, \apj, 755,
  138

\bibitem[{{Herant} {et~al.}(1992){Herant}, {Benz}, \& {Colgate}}]{Hera92}
{Herant}, M., {Benz}, W., \& {Colgate}, S. 1992, \apj, 395, 642

\bibitem[{{Herant} {et~al.}(1994){Herant}, {Benz}, {Hix}, {Fryer}, \&
  {Colgate}}]{Hera94}
{Herant}, M., {Benz}, W., {Hix}, W.~R., {Fryer}, C.~L., \& {Colgate}, S.~A.
  1994, \apj, 435, 339

\bibitem[{{Iwakami} {et~al.}(2008){Iwakami}, {Kotake}, {Ohnishi}, {Yamada}, \&
  {Sawada}}]{Iwak08}
{Iwakami}, W., {Kotake}, K., {Ohnishi}, N., {Yamada}, S., \& {Sawada}, K. 2008,
  \apj, 678, 1207

\bibitem[{{Janka} \& {Keil}(1998)}]{Jank98}
{Janka}, H.-T., \& {Keil}, W. 1998, in Supernovae and cosmology, ed.
  L.~{Labhardt}, B.~{Binggeli}, \& R.~{Buser}, 7

\bibitem[{{Janka} \& {Mueller}(1996)}]{Jank96}
{Janka}, H.-T., \& {Mueller}, E. 1996, \aap, 306, 167

\bibitem[{{Kraichnan}(1967)}]{Krai67}
{Kraichnan}, R.~H. 1967, Physics of Fluids, 10, 1417

\bibitem[{{Kuroda} {et~al.}(2012){Kuroda}, {Kotake}, \& {Takiwaki}}]{Kuro12}
{Kuroda}, T., {Kotake}, K., \& {Takiwaki}, T. 2012, \apj, 755, 11

\bibitem[{{Liebend{\"o}rfer} {et~al.}(2005){Liebend{\"o}rfer}, {Rampp},
  {Janka}, \& {Mezzacappa}}]{Lieb05}
{Liebend{\"o}rfer}, M., {Rampp}, M., {Janka}, H.-T., \& {Mezzacappa}, A. 2005,
  \apj, 620, 840

\bibitem[{{Marek} \& {Janka}(2009)}]{Mare09}
{Marek}, A., \& {Janka}, H.-T. 2009, \apj, 694, 664

\bibitem[{{Meakin} \& {Arnett}(2007)}]{Meak07}
{Meakin}, C.~A., \& {Arnett}, D. 2007, \apj, 667, 448

\bibitem[{{M{\"u}ller} {et~al.}(2012){M{\"u}ller}, {Janka}, \&
  {Marek}}]{Mull12}
{M{\"u}ller}, B., {Janka}, H.-T., \& {Marek}, A. 2012, \apj, 756, 84

\bibitem[{{Murphy} \& {Burrows}(2008)}]{Murp08}
{Murphy}, J.~W., \& {Burrows}, A. 2008, \apj, 688, 1159

\bibitem[{{Murphy} {et~al.}(2012){Murphy}, {Dolence}, \& {Burrows}}]{Murp12}
{Murphy}, J.~W., {Dolence}, J.~C., \& {Burrows}, A. 2012, ArXiv e-prints

\bibitem[{{Murphy} \& {Meakin}(2011)}]{Murp11}
{Murphy}, J.~W., \& {Meakin}, C. 2011, \apj, 742, 74

\bibitem[{{Nordhaus} {et~al.}(2010){Nordhaus}, {Burrows}, {Almgren}, \&
  {Bell}}]{Nord10}
{Nordhaus}, J., {Burrows}, A., {Almgren}, A., \& {Bell}, J. 2010, \apj, 720,
  694

\bibitem[{{Ott} {et~al.}(2008){Ott}, {Burrows}, {Dessart}, \& {Livne}}]{Ott08}
{Ott}, C.~D., {Burrows}, A., {Dessart}, L., \& {Livne}, E. 2008, \apj, 685,
  1069

\bibitem[{{Pejcha} \& {Thompson}(2012)}]{Pejc12}
{Pejcha}, O., \& {Thompson}, T.~A. 2012, \apj, 746, 106

\bibitem[{{Scheck} {et~al.}(2008){Scheck}, {Janka}, {Foglizzo}, \&
  {Kifonidis}}]{Sche08}
{Scheck}, L., {Janka}, H.-T., {Foglizzo}, T., \& {Kifonidis}, K. 2008, \aap,
  477, 931

\bibitem[{{Shen} {et~al.}(1998{\natexlab{a}}){Shen}, {Toki}, {Oyamatsu}, \&
  {Sumiyoshi}}]{Shen98a}
{Shen}, H., {Toki}, H., {Oyamatsu}, K., \& {Sumiyoshi}, K. 1998{\natexlab{a}},
  Nuclear Physics A, 637, 435

\bibitem[{{Shen} {et~al.}(1998{\natexlab{b}}){Shen}, {Toki}, {Oyamatsu}, \&
  {Sumiyoshi}}]{Shen98b}
---. 1998{\natexlab{b}}, Progress of Theoretical Physics, 100, 1013

\bibitem[{{Takiwaki} {et~al.}(2012){Takiwaki}, {Kotake}, \& {Suwa}}]{Taki12}
{Takiwaki}, T., {Kotake}, K., \& {Suwa}, Y. 2012, \apj, 749, 98

\bibitem[{{Thompson}(2000)}]{Thom00a}
{Thompson}, C. 2000, \apj, 534, 915

\bibitem[{{Thompson} {et~al.}(2003){Thompson}, {Burrows}, \& {Pinto}}]{Thom03}
{Thompson}, T.~A., {Burrows}, A., \& {Pinto}, P.~A. 2003, \apj, 592, 434

\bibitem[{{Woosley} \& {Weaver}(1995)}]{Woos95}
{Woosley}, S.~E., \& {Weaver}, T.~A. 1995, \apjs, 101, 181

\bibitem[{{Yamamoto} {et~al.}(2012){Yamamoto}, {Fujimoto}, {Nagakura}, \&
  {Yamada}}]{Yama12}
{Yamamoto}, Y., {Fujimoto}, S.-i., {Nagakura}, H., \& {Yamada}, S. 2012, ArXiv
  e-prints

\end{thebibliography}

\end{document}